\documentclass[sn-mathphys-num,icol]{sn-jnl}

\usepackage{graphics}
\usepackage{dcolumn}
\usepackage{bm}

\usepackage{array}
\usepackage{booktabs}
\usepackage{amsmath}
\usepackage{amssymb}
\usepackage{multirow}
\usepackage{xcolor}
\usepackage{graphicx}
\usepackage{epstopdf}
\usepackage[version=4]{mhchem}
\usepackage{verbatim}
\usepackage{subcaption}
\usepackage{fancyvrb}
\usepackage[export]{adjustbox}
\usepackage{svg} 
\usepackage{comment} 

\usepackage{float}
\usepackage{algorithm} 
\usepackage{algorithmic} 

\begin{document}

\title{DiffGA: Steering Frozen Diffusion Models for Materials Design via Evolutionary Search}

\author*[1]{Shivang Agarwal}\email{shivang.agarwal@sandboxaq.com}
\author[1]{Rodrigo Wang}
\author[1]{Kevin Ryczko}
\author[1]{Aayush R. Singh}
\affil[1]{SandboxAQ, Palo Alto, CA 94301, USA}

\abstract{Steering candidate generation pipelines toward user‑defined objectives is central to inverse materials design, yet it remains difficult when objectives are black‑box, ever-changing, or non‑differentiable. This is precisely the regime of traditional simulation‑ and experiment‑derived properties. Generative models offer a promising solution but have to be either retrained or fine‑tuned for every new objective, demanding large labeled datasets and substantial compute, or rely on differentiable score guidance that is brittle to gradient noise and undefined for many practical targets. We present DiffGA, a retraining‑free framework that injects evolutionary search directly into the sampling trajectory of a pretrained diffusion model, evolving partially denoised \textit{horizon} states under an arbitrary black‑box oracle. Often campaign cost is dominated by oracle evaluation and, thus, we couple DiffGA with an \textit{on‑the‑fly} Bayesian surrogate that reuses learned representations from state-of-the-art machine-learned force fields (MLFFs). A Gaussian process is fit during the campaign on the force field's own graph embeddings and --- once it demonstrably ranks an unseen generation --- replaces the expensive property‑evaluation step under an uncertainty‑aware acquisition rule, cutting oracle cost with negligible loss of accuracy. Across three crystal‑design tasks using a frozen MatterGen backbone, DiffGA outperforms diffusion‑only, compute‑matched, purely evolutionary, and fine‑tuned baselines in the (oracle‑assessed) rate of stable, unique, and novel hits. Rather than proposing specific new compounds, we use density functional theory to quantify the fidelity of the MLFF oracle itself, showing that it is a faithful proxy in key regimes and systematically biased in others, delineating precisely when proxy‑driven generative discovery can be trusted. On a representative sample, DiffGA's top hits survive first‑principles re‑evaluation as genuine, physically realistic hits, and the surrogate‑accelerated pipeline preserves this fidelity end‑to‑end at a fraction of the cost.}


\maketitle

\section*{Introduction}
\label{sec:intro}
Generative models have become powerful tools for designing complex, structured objects across images \cite{yu2023scaling, chang2023muse}, proteins \cite{ingraham2019generative, anand2018generative}, symbolic programs \cite{dinu2024symbolicai}, and scientific matter \cite{manica2023accelerating, sanchez2018inverse}. Among them, diffusion models stand out for producing high-fidelity, diverse samples through a denoising process \cite{zeni2025generative, sohl2015deep, ho2020denoising}. However, a central challenge persists which limits the widespread adoption and applicability of such technologies: how to steer a pretrained generative model toward \textit{user-defined} objectives. Achieving property- or goal-conditioned generation typically requires expensive fine-tuning on labeled data \cite{pyzer2025foundation, devi2024optimal, vlassis2023denoising, zeni2025generative} or architectural modifications for conditional guidance \cite{dhariwal2021diffusion, wang2024evaluating}. Gradient-based score guidance, the main alternative, assumes objectives that are differentiable, cheap to compute, and well defined on noisy intermediate states --- necessitating additional training on noisy inputs \cite{karras2022elucidating, xu2022geodiff}. In many real settings the objective is black-box, expensive, non-differentiable, or available only through simulations or experiments. The dynamic and evolving nature of such objectives particularly makes conventional diffusion guidance difficult or infeasible.

This challenge is acutely felt in materials discovery. Diffusion-based models have made rapid progress in generating valid, stable, and novel molecules and crystals \cite{zeni2025generative, jiao2023crystal, yang2023scalable}, most notably MatterGen \cite{zeni2025generative}, but they focus largely on unconditional generation or simple conditional tags such as composition, symmetry, or a single scalar property. Conditioning on multiple or bespoke properties, often computed via ever-changing computational workflows, requires assembling exhaustive labeled datasets and retraining the model for every new objective \cite{pyzer2025foundation, devi2024optimal, vlassis2023denoising}. Even then, such models cannot ingest on-the-fly, black-box feedback during sampling. Score-based guidance frameworks and fine-tuned models for materials and molecules \cite{kadan2025guided, schneuing2024structure, song2020score, hsu2024score} inherit the differentiability requirement and are typically tied to particular property predictors and representations.

Genetic algorithms (GAs) \cite{holland1992genetic} offer a complementary strength by performing a direct search over candidates through explicit selection, mutation, and recombination. They natively support black-box scoring and multi-objective optimization, and remain a staple of computational materials and molecular discovery \cite{ryczko2020inverse, tetsassi2021neural, kadan2023accelerated, henault2020chemical}. However, they construct candidates through stochastic mutation and crossover, which can produce chemically or structurally implausible configurations and consequently require explicit structural constraints, relaxation, or stability screening \cite{lee2023creation, yang2022crystal}. A growing body of work couples generative diffusion with evolutionary search, motivated by the formal correspondence between denoising and evolutionary selection \citep{zhang2025diffusion} and realized for images \citep{jajal2025inference} and, most closely, for molecules where genetic operators are applied to partially denoised states of a generator under repeated renoising-denoising loops \citep{sun2025diffusion}. In the materials setting, however, such coupling remains largely unexplored. Existing hybrids either apply genetic operators \textit{post hoc} to fully generated samples or rely on differentiable/learned surrogates tightly coupled to a specific representation \cite{ajagekar2025computer, gubina2024hybrid, nigam2022parallel, kneiding2024augmenting}. To our knowledge, no prior method evolves the intermediate diffusion states or the diffusion trajectory of a frozen, periodic-crystal generator such as MatterGen under an arbitrary black-box property oracle in a retraining-free framework.

\begin{figure*}[htb]
\centering
\includegraphics[width=\textwidth]{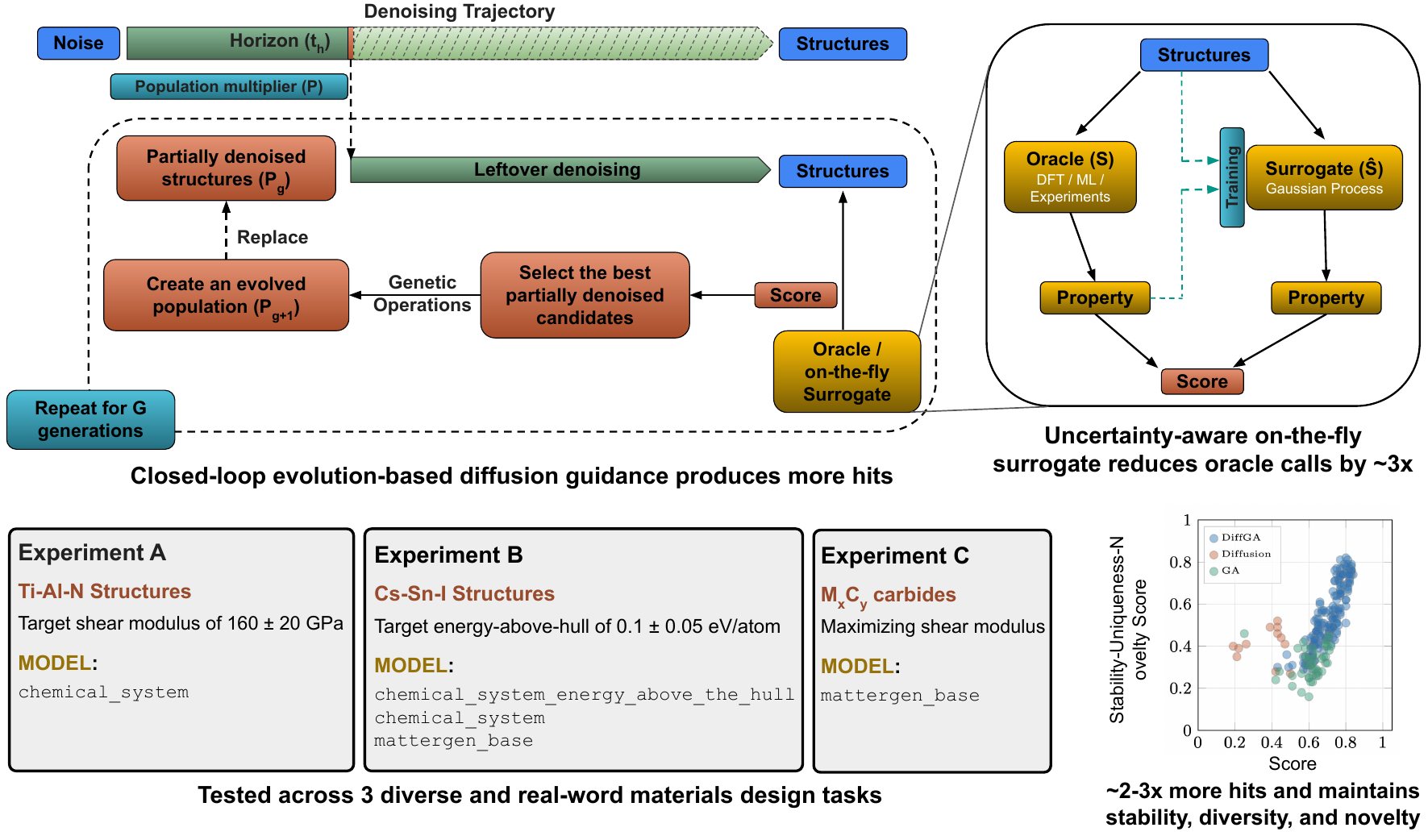}
\caption{A schematic overview of the DiffGA workflow for diffusion models and the experiments performed in this study. Details about each step can be found in the Methods section.}
\label{fig:diffga}
\end{figure*}

We introduce DiffGA, a hybrid framework that injects evolutionary search \textit{within} the diffusion sampling trajectory (Fig.~\ref{fig:diffga}). Rather than operating on finished samples, DiffGA acts on intermediate diffusion states, i.e., partially denoised candidates that capture the model's evolving generative hypotheses. Each intermediate state is completed through the remaining denoising steps, evaluated by an arbitrary (possibly black-box) oracle, and assigned a corresponding score via a simple scoring function. The scores drive a genetic selection-and-variation step over the intermediate states, producing the next population. By evolving the trajectory rather than the endpoint, DiffGA provides a flexible, gradient-free mechanism that both guides generation toward high-value regions of latent space and expands that space, while retaining the expressive power of the pretrained prior. Such a framework also avoids any additional retraining or fine-tuning, and objectives introduced entirely at inference time.

In materials discovery, the oracle is often the dominant cost. Thus, for all practical purposes, any framework built with the goal of materials discovery must be (i) flexible enough to accept any oracle with minimal changes to the rest of the workflow and (ii) efficient enough to afford the oracle at campaign scale. Specifically, for DiffGA, each candidate must be relaxed with a machine-learned force field (MLFF) and evaluated for stability and target properties, and this cost multiplies across every independent campaign. We therefore introduce a second contribution: an on-the-fly surrogate model that is trained \textit{during} the DiffGA run on the oracle evaluations it has already produced until that point. Rather than training a bespoke crystal-graph encoder \cite{xie2018crystal} or a deep-kernel network \cite{wilson2016deep} from scratch --- which is unstable in the low-data, continually shifting regime of a live campaign --- the surrogate reuses the representation the MLFF has already computed. From each relaxed candidate, we take the force field's own graph embedding and fit only a lightweight Gaussian process \cite{williams1995gaussian} that maps it to the oracle property, yielding predictive uncertainty. The surrogate is promoted to replace the expensive oracle property evaluation only after it demonstrably ranks an unseen generation which effectively acts as a prospective rank-correlation gate. It then scores candidates under an uncertainty-aware acquisition rule and is continually re-checked against the true oracle to prevent drift. This converts DiffGA from a method that tolerates expensive black-box objectives into one that makes black-box discovery tractable at large-scale discovery campaigns.

Finally, because a capable optimizer will exploit any weakness in its oracle, we do not rest our claims on MLFF-based oracles alone. However, we also do not aim to propose and fully characterize specific new compounds either. Instead, within the scope of this study, we use Density Functional Theory (DFT) to quantify the fidelity of the MLFF oracle itself, establishing (i) the regimes in which MLFF-evaluated properties and stability survive first‑principles re‑evaluation and those in which they are systematically biased, (ii) that the on‑the‑fly surrogate faithfully preserves search outcomes, and (iii) that DiffGA's top hits, on a representative sample, are genuine high‑property structures rather than artifacts of proxy optimization. This turns the central risk of proxy‑driven discovery into a measured quantity. Thus, the value of DiffGA is realized to the extent the oracle is trustworthy, which we make explicit.

\section*{Results}
\label{sec:results}

\subsection*{DiffGA steers frozen generative models without training or fine-tuning}
\label{subsec:base_results}
We evaluate DiffGA on three representative materials-design tasks selected to probe complementary aspects of oracle-guided generation: targeted optimization within a constrained chemical space, optimization when a relevant property-conditioned generative prior is already available, and open-ended optimization in a broader chemical space where such conditioning is difficult to obtain. Together, these tasks progressively increase the demands on the search while testing whether oracle-guided evolutionary feedback provides benefits beyond those of the underlying generative prior.
\begin{enumerate}
    \item \textbf{Experiment A}: Generate Ti-Al-N structures with a target shear modulus of $160 \pm 20$ GPa. This encompasses a narrow chemical system with a fixed property target.
    \item \textbf{Experiment B}: Generate Cs-Sn-I structures with a target energy above the hull of $0.1 \pm 0.05$ eV/atom. The target here is directly supported by an existing property-conditioned MatterGen model, enabling ablation-like comparisons with fine-tuned models and helping isolate the effect of DiffGA.
    \item \textbf{Experiment C}: Maximize the shear modulus of transition-metal carbides, $\mathrm{M}_x\mathrm{C}_y$, where $\mathrm{M}\in [\mathrm{Ti}, \mathrm{V}, \mathrm{Cr},\mathrm{Mo},\mathrm{W}]$. This is an open-ended search space where fine-tuning is highly non-trivial.
\end{enumerate}

All experiments use a frozen MatterGen backbone from one of three pretrained variants --- \texttt{mattergen\_base} (unconditioned), \texttt{chemical\_system} (conditioned on chemical system), and \texttt{chemical\_system\_energy\_above\_hull} (conditioned on chemical system and energy above the hull) --- with no retraining or fine-tuning of the generative model. Across all tasks the framework is unchanged except for the oracle and scoring function being used (details can be found under the Scoring function and Oracle sections in Methods), isolating the effect of the DiffGA mechanism from that of model-specific conditioning. We use five horizon times ($t_h$) --- the intermediate timestep where denoising is stopped --- from $\{0.98, 0.75, 0.5, 0.25, 0.002\}$ and three population multipliers $P \in \{1, 2, 3\}$, repeating every configuration over three random seeds. For experiments A and B, we set the number of generations to $G=8$. This is increased to $G=15$ whenever the \texttt{mattergen\_base} model is used in experiments B and C to allow more iterations for optimizing both the chemical space and the material property. Batch size is maintained at $B=32$ throughout. Further details can be found in the Methods section. The limit $t_h = 0.002$ corresponds to a single remaining denoising step per generation and thus serves as a pure-GA baseline, as discussed in detail in section S1 of the Supplementary Information (SI). While the stochastic nature of generative sampling and the use of three random seeds may limit the statistical power of fine-grained comparisons, we argue that each `seed' comprises of $15$ individual runs using different configurations and that our conclusions therefore focus on consistent trends across tasks, horizon times, and population sizes.

For every experiment, a structure is classified as a `hit' if it satisfies the task's property constraints up to a predefined threshold. This suffices for experiments A and B. For experiment C, since there is no target property value, a candidate structure is classified as a hit if it has a shear modulus greater than or equal to $95\%$ of the maximum value found in the Materials Project (MP) database \citep{jain2013commentary} for that chemical space, suggesting a search outside the typical training regime. The search space for experiment C has been restricted to the set of $5$ transition metals because their respective carbides are the canonical high-shear-modulus refractory hard materials targeted by the objective, whereas later transition metals form comparatively soft, often metastable carbides that lie outside the high-modulus regime under study \cite{lengauer2000transition, jhi1999electronic, toth1971transition, pierson1996handbook}. Lastly, following MatterGen, we additionally report SUN (Stable, Unique, Novel) hit rates, which measure the physical validity and diversity of successful candidates. Throughout, SUN rates are computed by the oracle, i.e., MatterSim‑relaxed energy above the hull for stability \citep{yang2024mattersim, zeni2025generative}, and \texttt{pymatgen}'s \texttt{StructureMatcher} \citep{ong2013python} for uniqueness and novelty against the reference set, and therefore serve as a common‑evaluator measure of physical validity and diversity for comparing methods, not as a claim that individual hits are experimentally novel compounds. First‑principles validation of the oracle underlying these metrics is treated separately (see First-principles validation and oracle fidelity in the Results section).

\begin{table*}[htbp]
\caption{Hit rate (HR) metrics for each experiment. We compare diffusion (first-generation\textsuperscript{$\dagger$} and compute-matched baselines\textsuperscript{$\ddagger$}), GA, and DiffGA, and evaluate performance across three pretrained MatterGen variants: \texttt{mattergen\_base}\textsuperscript{a}, \texttt{chemical\_system}\textsuperscript{b}, and \texttt{chemical\_system\_energy\_above\_hull}\textsuperscript{c}.}
\label{tab:hit_rates_and_scoring_funcs}
\begin{small}
\begin{tabular*}{\textwidth}{@{\extracolsep\fill}lcccr}
\toprule
Exp. & Method & Mean HR (\%) & Mean SUN HR (\%) & Best SUN HR (\%) \\
\midrule
\multirow{4}{*}{A} & Diff.\textsuperscript{$\dagger$} & 14.51\textsuperscript{\normalfont b} & 1.38\textsuperscript{\normalfont b} & 4.17\textsuperscript{\normalfont b} \\
         & Diff.\textsuperscript{$\ddagger$} & 82.98\textsuperscript{\normalfont b} & 8.18\textsuperscript{\normalfont b} & 15.78\textsuperscript{\normalfont b} \\
         & GA & 77.78\textsuperscript{\normalfont b} & 8.58\textsuperscript{\normalfont b} & \textbf{19.22}\textsuperscript{\normalfont b} \\
         & DiffGA & \textbf{83.26}\textsuperscript{\normalfont b} & \textbf{9.55}\textsuperscript{\normalfont b} & 19.22\textsuperscript{\normalfont b} \\
\midrule
\multirow{4}{*}{B} & Diff.\textsuperscript{$\dagger$} & 2.08\textsuperscript{\normalfont c} / 1.46\textsuperscript{\normalfont b} / 0.07\textsuperscript{\normalfont a} & 0.58\textsuperscript{\normalfont c} / 0.19\textsuperscript{\normalfont b} / 0.02\textsuperscript{\normalfont a} & 2.08\textsuperscript{\normalfont c} / 0.69\textsuperscript{\normalfont b} / 0.35\textsuperscript{\normalfont a} \\
         & Diff.\textsuperscript{$\ddagger$} & 3.47\textsuperscript{\normalfont c} / 1.74\textsuperscript{\normalfont b} / 0.11\textsuperscript{\normalfont a} & 1.73
         \textsuperscript{\normalfont c} / 0.69\textsuperscript{\normalfont b} / 0.08\textsuperscript{\normalfont a} & 9.37\textsuperscript{\normalfont c} / 6.25\textsuperscript{\normalfont b} /0.42\textsuperscript{\normalfont a} \\
         & GA & \textbf{100}\textsuperscript{\normalfont c} / \textbf{92.36}\textsuperscript{\normalfont b} / 23.26\textsuperscript{\normalfont a} & 44.32\textsuperscript{\normalfont c} / 38.76\textsuperscript{\normalfont b} / 17.82\textsuperscript{\normalfont a} & 57.29\textsuperscript{\normalfont c} / 49.60\textsuperscript{\normalfont b} / 23.15\textsuperscript{\normalfont a} \\
         & DiffGA & 90\textsuperscript{\normalfont c} / 84.03\textsuperscript{\normalfont b} / \textbf{50.56}\textsuperscript{\normalfont a} & \textbf{46.02}\textsuperscript{\normalfont c} / \textbf{39.46}\textsuperscript{\normalfont b} / \textbf{42.54}\textsuperscript{\normalfont a} & \textbf{72.92}\textsuperscript{\normalfont c} / \textbf{70.83}\textsuperscript{\normalfont b} / \textbf{71.97}\textsuperscript{\normalfont a} \\
\midrule
\multirow{4}{*}{C} & Diff.\textsuperscript{$\dagger$} & 0\textsuperscript{\normalfont a} & 0\textsuperscript{\normalfont a} & 0\textsuperscript{\normalfont a} \\
         & Diff.\textsuperscript{$\ddagger$} & 0\textsuperscript{\normalfont a} & 0\textsuperscript{\normalfont a} & 0\textsuperscript{\normalfont a} \\
         & GA & 10.07\textsuperscript{\normalfont a} & 0.17\textsuperscript{\normalfont a} & 0.52\textsuperscript{\normalfont a} \\
         & DiffGA & \textbf{12.85}\textsuperscript{\normalfont a} & \textbf{0.58}\textsuperscript{\normalfont a} & \textbf{2.19}\textsuperscript{\normalfont a} \\
\bottomrule
\end{tabular*}
\end{small}
\end{table*}

We compare against four baselines, details for which can be found in section S1 of the SI:
\begin{enumerate}
    \item Diffusion baseline ($\text{Diff.}^\dagger$): sampling $B$ structures directly from the generative model, identical to the output after the first generation of DiffGA.
    \item Compute-matched diffusion ($\text{Diff.}^\ddagger$): sampling $P \times B \times G$ structures and ranking them by the same oracle, isolating the effect of oversampling and oracle access without iterative feedback.
    \item GA: identical to a DiffGA run but with the horizon time, $t_h$, set to $0.002$.
    \item Sampling using fine-tuned domain-specific MatterGen variants, wherever available and applicable.
\end{enumerate}

Across all three experiments, DiffGA achieves the highest SUN hit rates while remaining competitive or superior on overall hit rate (see Table \ref{tab:hit_rates_and_scoring_funcs}). To make the effect of DiffGA more explicit, we report mean hit rates as averaged across all seeds and hyperparameters as well as the `best' hit rate across them. Only experiment B is directly supported by an existing property-conditioned MatterGen model, and experiment A becomes accessible only through additional fine-tuning. Experiment C is out of reach for any currently available diffusion model, yet DiffGA succeeds using only the unconditioned prior. Notably, the standalone \texttt{chemical\_system\_energy\_above\_hull} model under the diffusion baselines performs poorly on Experiment B when compared to DiffGA, plausibly because the Cs-Sn-I space was poorly covered in its fine-tuning set --- a concrete illustration of the fragility of retraining-based approaches, which DiffGA sidesteps.

While GA can match DiffGA on the overall hit rate (Experiment B), DiffGA consistently yields higher SUN-consistent hit rates, i.e., more successful structures that are simultaneously stable, unique, and novel. GA tends to concentrate on a small set of high-scoring motifs, whereas DiffGA preserves structural diversity through the generative prior while applying evolutionary pressure. The advantage is most pronounced with unconditioned models, where GA lacks any structural bias. For experiment C, DiffGA outperforms GA by factors of $\approx3.5$ and $\approx4$ in mean and best SUN hit rate, respectively. Given the extreme sparsity of viable solutions in the open-ended carbide space and the use of an unconditioned prior, this gap is direct evidence for the value of combining evolutionary search with a generative prior over purely evolutionary search. Lastly, the compute-matched baseline results showcase that iterative oracle feedback provides greater optimization efficiency than post-hoc ranking of an equal number of independent diffusion samples.

\subsection*{Evolutionary feedback compounds over generations}
\label{subsec:hit_rates_results}
\begin{figure*}[htbp]
\centering
\subfloat[$P=1$]{\includegraphics[width=0.31\linewidth]{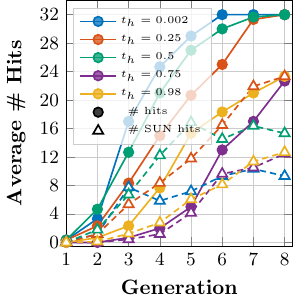}}
\hfill
\subfloat[$P=2$]{\includegraphics[width=0.31\linewidth]{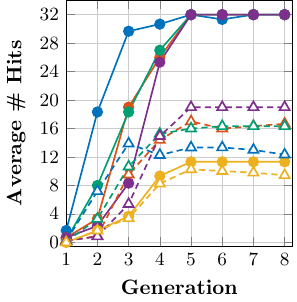}}
\hfill
\subfloat[$P=3$]{\includegraphics[width=0.31\linewidth]{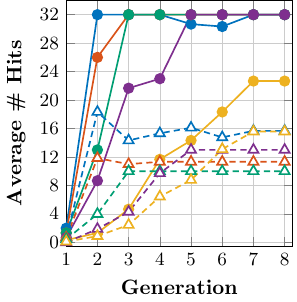}}
\vspace{0.75cm}
\subfloat[$P=1$]{\includegraphics[width=0.31\linewidth]{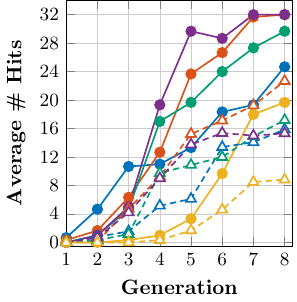}}
\hfill
\subfloat[$P=2$]{\includegraphics[width=0.31\linewidth]{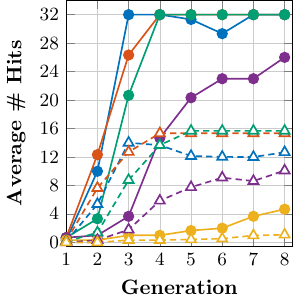}}
\hfill
\subfloat[$P=3$]{\includegraphics[width=0.31\linewidth]{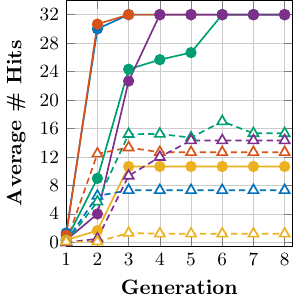}}
\vspace{0.75cm}
\subfloat[$P=1$]{\includegraphics[width=0.31\linewidth]{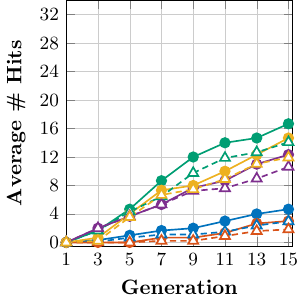}}
\hfill
\subfloat[$P=2$]{\includegraphics[width=0.31\linewidth]{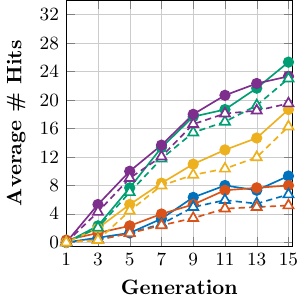}}
\hfill
\subfloat[$P=3$]{\includegraphics[width=0.31\linewidth]{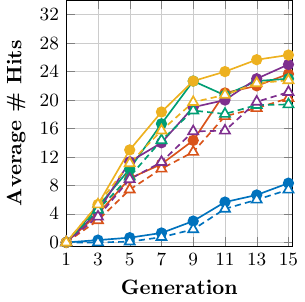}}
\caption{The average number of hits across generations for experiment B for each value of $P$, averaged across the three seeds. The models used are: (a, b, c) \texttt{chemical\_system\_energy\_above\_hull}, (d, e, f) \texttt{chemical\_system}, and (g, h, i) \texttt{mattergen\_base}. For clarity, the legend has only been shown in the first plot but applies to all figures, and for (g, h, i), we plot every $2^{\text{nd}}$ data point.}
\label{fig:hits_experiment_B}
\end{figure*}

To elucidate the mechanism behind these differences, we track the mean number of hits and SUN hits across generations (Fig.~\ref{fig:hits_experiment_B}) for experiment B. Similar plots for experiments A and C can be found in section S3 of the SI. Across all model variants, DiffGA shows a clear upward trend in both metrics, indicating that evolutionary feedback steadily increases both the likelihood of satisfying the target and the number of distinct, stable solutions discovered. In contrast, the GA limit ($t_h = 0.002$) can reach competitive raw hit counts but consistently underperforms in SUN hits, reflecting reduced structural diversity among its successes. Occasional dips in hit counts across generations are artifacts of elitist truncation (the retained set is bounded by top-$k$ selection) and of early stopping averaged across seeds, not losses in solution quality. This is explained by the observation that the top-quartile score rises monotonically regardless, as seen in the detailed evolution of scores and properties across generation for different values of $t_h$ and $P$ for each experiment (see sections S2 and S3 in the SI).

\subsection*{DiffGA steers the frozen prior into a target region of its own latent space}
\label{subsec:umap_embedding}
\begin{figure}[htbp]
\centering
\includegraphics[width=0.5\linewidth]{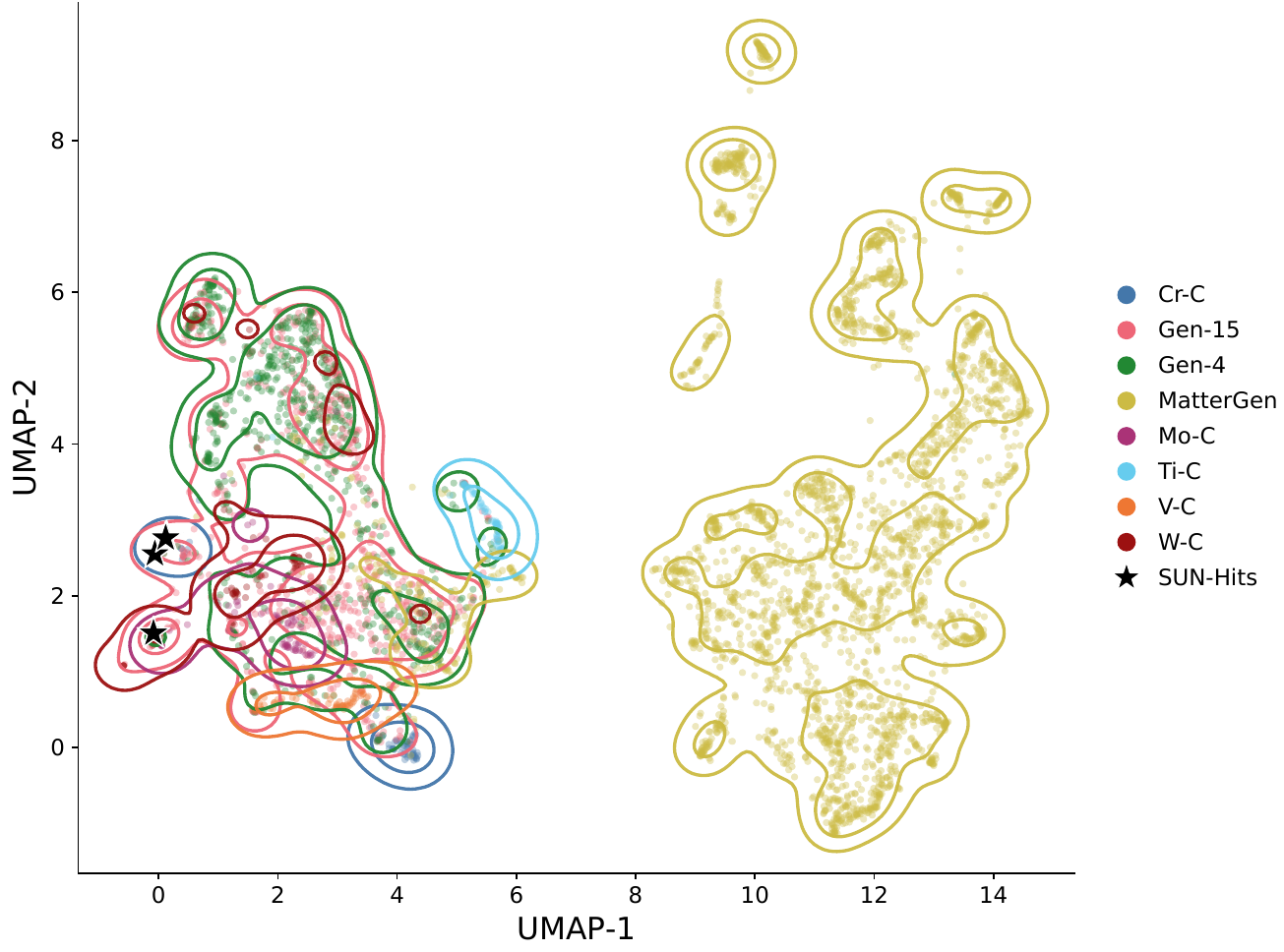}
\caption{Joint UMAP of pooled GemNet node embeddings evaluated at the final denoising step for unconditional \texttt{mattergen\_base} samples, reference transition metal carbides, and DiffGA populations at generations 4 and 15 for experiment C. Contours enclose 50\% and 90\% of each set's density. The black stars mark the four DFT‑validated high‑shear‑modulus hits (Table \ref{tab:dft_tmc_sun_hits}).}
\label{fig:embedding}
\end{figure}

To see where in the generative model's own representation DiffGA directs its search, we visualize MatterGen's learned embedding space directly (Fig.~\ref{fig:embedding}) in the context of experiment C. We consider the following sets of structures: (i) every fully denoised, relaxed structure from the set of unconditional samples from the frozen \texttt{mattergen\_base} prior, (ii) reference carbides for each of the five chemistries (generated with the \texttt{chemical\_system} model together with Materials Project entries), and (iii) DiffGA's populations at generations 4 --- when the expensive oracle is switched on (see Computational cost in the Methods section) --- and 15 --- the last generation. We then extract the per‑atom node embeddings of MatterGen's GemNet \citep{gasteiger2021gemnet} backbone at the final denoising step, mean‑pool them into a single 512‑dimensional descriptor, and project all descriptors jointly with one UMAP \citep{mcinnes2018umap}. Because every structure is embedded by the same frozen model at the same timestep and placed with a single shared projection, the sets are directly comparable. As expected, the unconditional samples spread broadly, reflecting the prior's chemical diversity, whereas the transition metal carbides occupy a distinct, compact region of the same space. DiffGA's populations sit squarely within this carbide region. In fact, it is easy to observe that by generation 4 the structural embeddings have already concentrated heavily around the embedding space of interest and increase their overlap with this space significantly by generation 15. This shows that, with no retraining, DiffGA progressively concentrates generated structures in a region of the underlying diffusion model's representation space enriched in the structures of interest (here, transition-metal carbides structures with high shear moduli). The four DFT‑validated high‑shear‑modulus SUN hits (Table \ref{tab:dft_tmc_sun_hits}) lie inside this region, confirming that the sub‑region DiffGA selects contains genuinely stiff, physically realistic structures rather than proxy artifacts.

\subsection*{DiffGA improves the score-diversity trade-off}
\label{subsec:pareto_results}
\begin{figure*}[htbp]
\centering
\subfloat[]{\includegraphics[width=0.31\linewidth]{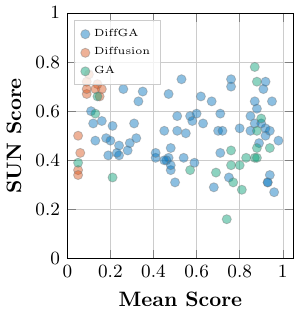}}
\hfill
\subfloat[]{\includegraphics[width=0.31\linewidth]{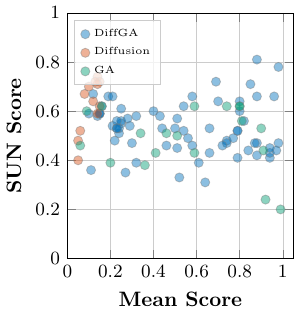}}
\hfill
\subfloat[]{\includegraphics[width=0.31\linewidth]{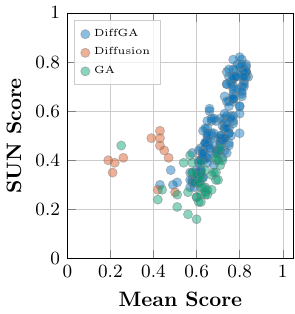}}
\caption{Pareto frontier plots for experiment B using different models: (a) \texttt{chemical\_system\_energy\_above\_hull}, (b) \texttt{chemical\_system}, and (c) \texttt{mattergen\_base}.}
\label{fig:all_chemsys_hull_paretos}
\end{figure*}

To assess the trade-off between objective optimization and structural diversity, we examine Pareto frontiers over oracle scores and SUN metrics (Fig.~\ref{fig:all_chemsys_hull_paretos} for experiment B, Supplementary Fig.~S7 for experiments A and C). DiffGA consistently attains superior trade-offs --- higher scores without collapsing diversity --- relative to both GA and diffusion baselines. This supports the conclusion that DiffGA improves not only final performance metrics but the structure of the search process itself. It concentrates probability mass on high-value regions while retaining the generative prior's diversity, exactly the behavior needed for downstream discovery where a portfolio of viable, distinct candidates is more valuable than a single optimum.

\subsection*{An on-the-fly surrogate makes black-box campaigns tractable}
\label{subsec:surrogate_results}
The results thus far use the true oracle for every candidate. Because a campaign performs $N_\text{oracle} = P \times B \times G$ evaluations per seed (see Computational cost in the Methods section) and the property-evaluation stage can dominate that cost, we investigate how much of it can be avoided without changing outcomes. Consequently, we introduce an on-the-fly surrogate that reuses the representation the MLFF has already computed. From each relaxed structure, we take MatterSim's internal graph embedding and fit a Gaussian process \cite{williams1995gaussian} that maps it to the oracle property.

The design has three deliberate features, each addressing a failure mode we encountered. First, nothing is trained from scratch. A bespoke encoder based on crystal graph convolutional neural networks \cite{xie2018crystal} proved unstable in the shifting, low-data setting of a live run, so the only thing learned online is a small set of GP hyperparameters over a fixed, physically informative descriptor. Second, the surrogate is promoted only after it demonstrably ranks an unseen generation. This acts as a prospective Spearman-rank-correlation gate rather than a fit statistic on already-seen data, motivated by the fact that the GA consumes only the order, i.e., the ranking, of candidates, and rank quality (unlike $R^2$) does not collapse as the search concentrates the property variance. Third, once promoted, it delegates candidates to the scorer under an optimistic, uncertainty-aware acquisition, and a fraction of its most-uncertain predictions is re-checked against the true oracle each generation, so the search keeps exploring and the model is continually re-anchored. Most importantly, this is a drop-in modification and the DiffGA loop and objective remain unchanged. Further implementation details and Spearman rank correlation metrics across generations can be found in the Methods section where we show that the surrogate indeed maintains strong prospective ranking performance throughout the campaigns (Fig.~\ref{fig:bo_spearman_rank_corr}).

\begin{figure*}[htbp]
\centering
\subfloat[Experiment A]{\includegraphics[width=0.35\linewidth]{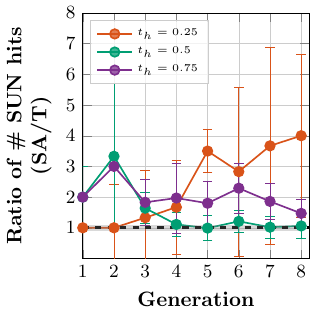}\label{subfig:sun_hits_ratio_expA}}
\hfill
\subfloat[Experiment B]{\includegraphics[width=0.295\linewidth]{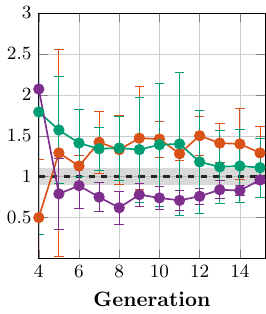}\label{subfig:sun_hits_ratio_expB}}
\hfill
\subfloat[Experiment C]{\includegraphics[width=0.295\linewidth]{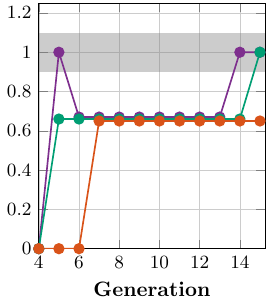}\label{subfig:sun_hits_ratio_expC}}
\vspace{0.75cm}
\subfloat[Experiment A]{\includegraphics[width=0.36\linewidth]{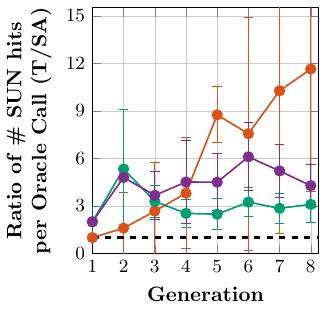}\label{subfig:sun_hits_ratio_per_compute_expA}}
\hfill
\subfloat[Experiment B]{\includegraphics[width=0.28\linewidth]{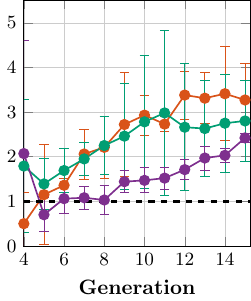}\label{subfig:sun_hits_ratio_per_compute_expB}}
\hfill
\subfloat[Experiment C]{\includegraphics[width=0.28\linewidth]{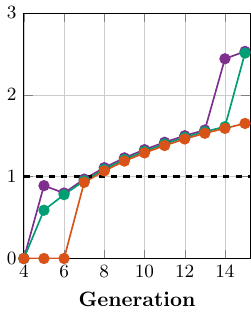}\label{subfig:sun_hits_ratio_per_compute_expC}}
\caption{(a, b, c) Optimization performance preservation and (d, e, f) oracle efficiency improvement for different horizon times, $t_h$, averaged over three seeds. For (a, b, c), the ratio shown is the number of SUN hits for the surrogate-accelerated (SA) pipeline to those for the traditional (T) DiffGA pipeline. For (d, e, f), the ratio is reversed. Experiment A used the \texttt{chemical\_system} model, whereas experiments B and C use the \texttt{mattergen\_base} model. For clarity, the y axes labels and legends have only been shown on the first column but hold true for all plots. For experiment C, error bars have been omitted due to large variations (courtesy of extremely low hit rates). Overall, the surrogate-accelerated pipeline maintains comparable performance while drastically increasing efficiency.}
\label{fig:bo_performance}
\end{figure*}

We evaluate the surrogate-accelerated pipeline on the same three tasks along two critical axes: (i) preservation of hit quality across generations, and (ii) reduction in the number of oracle calls. Thus, we compute both the number of SUN hits as well as the number of SUN hits per oracle call and, in Fig.~\ref{fig:bo_performance}, plot the ratio of these two quantities between the surrogate-accelerated (SA) pipeline and the traditional (T) DiffGA pipeline. To isolate the surrogate's effect from run-to-run sampling noise, both pipelines share an identical first-generation population and differ only in the scoring signal, i.e., true property or surrogate prediction under acquisition. To avoid the risk of overfitting, we set $B=128$ and repeat the experiments for all horizon times but set $P=1$ (note that this is similar, but not equivalent, to setting $B=32$ and $P=4$). We used $t_h\in\{0.25,0.5,0.75\}$ (note that $t_h=0.002$ and $t_h=0.98$ were simply used as baselines for previous analysis).

Figs.~\ref{subfig:sun_hits_ratio_expA}, \ref{subfig:sun_hits_ratio_expB}, \ref{subfig:sun_hits_ratio_expC} show that the surrogate‑accelerated (SA) pipeline retains the SUN‑hit trajectory of the traditional (T) pipeline across horizon times, while Figs.~\ref{subfig:sun_hits_ratio_per_compute_expA}, \ref{subfig:sun_hits_ratio_per_compute_expB}, \ref{subfig:sun_hits_ratio_per_compute_expC} show the corresponding gain in oracle efficiency, frequently exceeding a factor of three. The efficiency arises from two effects: (i) an actively trained, uncertainty‑aware surrogate ranks candidates well enough to replace the expensive property call, and (ii) once promoted, only a fraction $\rho=0.3$ of the most‑uncertain candidates are re‑queried against the true oracle each generation, so expensive property evaluations scale as $PB\times[1 + \rho(G-1)]$ rather than $PBG$. Because, for experiments B and C, the expensive property evaluator (oracle) is activated only from generation 4 onward (early generations steer the population into the target chemical space using inexpensive structural criteria; see Computational cost in the Methods section), their curves begin at that generation. Averaged over horizon times, the mean SUN hit rates for the traditional and surrogate‑accelerated pipelines are ($5.21$\%, $6.27$\%) for Experiment A, ($26.74\%$, $28.21$\%) for Experiment B, and ($0.60$\%, $0.43$\%) for Experiment C. Thus, accuracy is preserved or slightly improved for A and B, and statistically indistinguishable for C, where absolute counts are extremely low (1-2 SUN hits per batch of structures). We note here that because of a different configuration of inputs used to study the effect of the surrogate (namely, a larger batch size and fewer horizon times), these rates are not directly comparable to Table \ref{tab:hit_rates_and_scoring_funcs} but show the same qualitative preservation. Together, this establishes that the surrogate converts DiffGA into a campaign-scale tool at a fraction of the oracle budget, opening the door to costlier oracles such as higher‑fidelity simulations or experiments.

At this point, it is worth noting that the use of the surrogate is optional and, in some cases, may be redundant if the oracle itself is cheap. This might be the case when the oracle is a lightweight MLFF or an analytical function. However, as the oracle becomes costlier, such as for DFT-in-the-loop discovery campaigns, the surrogate can drastically accelerate the pipeline by reducing the number of oracle calls. The advantage of the on-the-fly surrogate, thus, is closely related to the true cost of the oracle.

\subsection*{First‑principles validation and oracle fidelity}
\label{subsec:dft_results}
Because DiffGA is an effective optimizer, its performance on MLFF‑evaluated objectives could, in principle, reflect exploitation of scorer or oracle weaknesses rather than genuine materials quality. We therefore validate a representative sample of hits with plane‑wave DFT \citep{car1985unified}, with the explicit goal of quantifying the fidelity of the oracle rather than proposing specific compounds.

Throughout the study, the shear modulus was calculated using MatTen \cite{wen2024equivariant} (see the Oracle section under Methods). Over a representative set of Ti-Al-N and transition metal carbide structures generated at the last generation for experiments A and C, MatTen-predicted shear moduli correlate strongly with DFT clamped-ion shear moduli (Pearson's r = $0.82$, Fig.~\ref{subfig:clamped_ion_dft}) but only weakly with DFT relaxed‑ion moduli (Pearson's r = $0.27$, Fig.~\ref{subfig:relaxed_ion_dft}). This interesting observation is instructive rather than indicative of a failure of DiffGA. The internal‑strain (ionic‑relaxation) contribution to the elastic tensor is negative‑semi-definite, so the relaxed‑ion modulus is bounded above by the clamped‑ion modulus. For the structures generated across experiments A and C, MatTen is found to be faithful to the clamped‑ion quantity it represents. The gap between the two is therefore a clean, physically interpretable characterization of the oracle's scope, not noise.

For structures generated through Experiment B, oracle- and DFT-computed energies above the hull have a Pearson correlation coefficient of $0.71$ (Fig.~\ref{subfig:ehull_dft}). We also note that this comparison is sensitive to functional consistency. An initial evaluation that placed PBE‑level \citep{ernzerhof1999assessment} MLFF energies against an r2SCAN \cite{furness2020accurate} DFT hull produced a spurious constant offset, leading to a Pearson correlation coefficient of $0.785$ but a $R^2$ value of $-0.25$. Analyzing the same candidates and the hull with a consistent functional (GGA/PBE for both) removes the offset, resulting in a Pearson correlation coefficient of $0.707$ and $R^2 = 0.17$. We report this explicitly because functional consistency between the surrogate/oracle and its DFT reference is a common and easily-overlooked source of apparent oracle ``error'' in generative‑discovery pipelines.

\begin{figure*}[htbp]
\centering
\subfloat[Relaxed-ion shear modulus]{\includegraphics[width=0.31\linewidth]{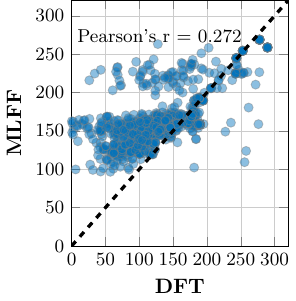}\label{subfig:relaxed_ion_dft}}
\hfill
\subfloat[Clamped-ion shear modulus]{\includegraphics[width=0.31\linewidth]{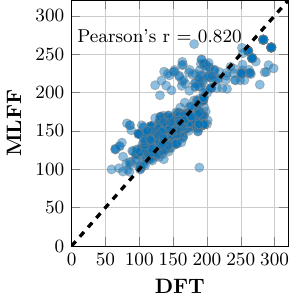}\label{subfig:clamped_ion_dft}}
\hfill
\subfloat[Energy above the hull]{\includegraphics[width=0.31\linewidth]{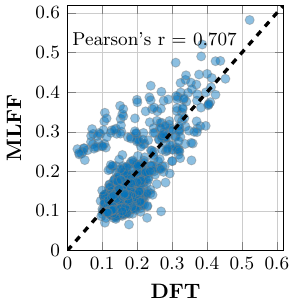}\label{subfig:ehull_dft}}
\caption{Oracle fidelity across experiments by comparing against DFT baselines. (a) Relaxed-ion shear moduli (GPa) and (b) clamped-ion shear moduli (GPa) correlation for Ti-Al-N and transition metal carbide systems, and (c) energy above the hull (eV/atom) correlation for Cs-Sn-I systems.}
\label{fig:dft_validation}
\end{figure*}

For the highest‑scoring SUN carbide hits, DFT calculations confirm genuinely high shear moduli, as can be seen in Table \ref{tab:dft_tmc_sun_hits}. The MLFF oracle in fact underestimates these moduli by $\approx$ 15-35 GPa, so the hits are, if anything, better than the oracle judged them. Critically, the DFT‑relaxed structures are essentially identical to the MLFF‑relaxed inputs (normalized RMSD $\leq4\times10^{-3}$), so the hits occupy the same region of structure space before and after first‑principles relaxation. Moreover, their energies above the hull, as computed by DFT, range from $0.01$ --- $0.12$ eV/atom. The WC hit sits comfortably within the $0.1$ eV/atom threshold as used in the MatterGen study as well as in our SUN calculations, while the three CrC polymorphs fall marginally above it, consistent with the oracle's mild optimism on stability quantified in Fig.~\ref{subfig:ehull_dft}. All four remain metastable, and their high shear moduli are confirmed regardless. These candidates include well-known ultra-hard carbides, which we take as evidence that DiffGA, starting from an unconditioned prior, correctly navigates to the genuinely stiff region of carbide space (Fig.~\ref{fig:embedding}). Once again, we note here that with these results, the aim is to show validation of the method, rather than a claim of new compounds.

\begin{table*}[htbp]
\caption{DFT validation of the top transition metal carbide SUN hits showing MatTen- and DFT-computed shear moduli (in GPa), normalized RMSD between the MLFF‑relaxed and DFT‑relaxed structures, and DFT (r2SCAN) energy above hull (eV/atom). The highest shear moduli (G) for a given chemical space reported in the MP database is also presented.}
\label{tab:dft_tmc_sun_hits}
\begin{small}
\begin{tabular*}{\textwidth}{@{\extracolsep\fill}lcccccr}
\toprule
SUN & MatTen & DFT & DFT & Highest MP & Normalized & Energy above the \\
Hit & (GPa) & Clamped-ion (GPa) & Relaxed-ion (GPa) & value (GPa) & RMSD & hull (eV/atom) \\
\midrule
WC & $261.18$ & $295.26$ & $289.58$ & $263.151$ & $1.6\times10^{-3}$ & $0.0164$ \\
CrC & $228.06$ & $263.73$ & $254.15$ & $249.439$ & $3.6\times10^{-3}$ & $0.1023$ \\
CrC & $236.41$ & $252.33$ & $243.01$ & $249.439$ & $1.9\times10^{-3}$ & $0.1255$ \\
CrC & $221.70$ & $251.18$ & $242.57$ & $249.439$ & $1.6\times10^{-4}$ & $0.1144$ \\
\bottomrule
\end{tabular*}
\end{small}
\end{table*}

Taken together, these checks show that DiffGA's advantage is realized in first‑principles‑validated materials quality wherever the oracle is faithful, and they map exactly where it is not. Thus, while the success of this framework is rooted heavily in the accuracy of the oracle being used, the highly flexible choice of oracles and scorers afforded by DiffGA allows drop-in replacements to be made with minimal changes elsewhere.

\section*{Conclusion}
\label{sec:discussion}
We have shown that evolutionary feedback applied to the intermediate states of a pretrained diffusion model enables effective, goal‑directed materials generation with no retraining or fine‑tuning, even for unconditioned or weakly conditioned priors. Across three tasks, DiffGA improves over diffusion‑only and purely evolutionary baselines in both overall and SUN‑consistent hit rates, and matches or exceeds domain‑specific fine‑tuned models --- succeeding even in the open‑ended transition‑metal‑carbide space, where fine‑tuning is impractical. The two ingredients are complementary rather than incremental: the generative prior supplies the structural validity and diversity that a pure genetic algorithm lacks, while evolutionary feedback supplies the exploitative, black‑box‑compatible steering that differentiable diffusion guidance cannot.

This retraining‑free design is what makes DiffGA practical when objectives are black‑box, expensive, non‑differentiable, or available only through simulation or experiment. Because the model is frozen, cost is incurred only at inference and a single prior is reused across every objective, making DiffGA more economical than retraining in the multi‑task, exploratory setting of discovery --- particularly when objectives change frequently or labeled datasets are unavailable. The on‑the‑fly surrogate compounds this advantage by reusing the force field's own representation and taking over the expensive property step once it can rank an unseen generation, removing much of the dominant cost term. Moreover, its benefit grows with oracle cost, making it most valuable precisely for the expensive simulation‑ or experiment‑level oracles that motivate the approach. The principal alternative --- reinforcement‑learning fine‑tuning of the generator --- instead bakes a target into the model weights, but requires a full fine‑tuning run per objective, is sensitive to reward‑model and KL‑regularization choices, and cannot ingest a genuinely black‑box oracle without a differentiable or learned reward. DiffGA trades this for inference‑time search, which is preferable in the multi-objective, black-box, frequently-changing-objective regime of exploratory discovery.

DiffGA sits within an emerging paradigm of evolutionary steering of diffusion models, in which the denoising trajectory itself is treated as a population to be optimized \citep{zhang2025diffusion}. It differs from the closest such method \citep{sun2025diffusion} in three key aspects: (i) it targets periodic inorganic crystals, (ii) it persistently evolves a horizon‑state population and truly steers the diffusion denoising trajectory rather than re‑noising completed samples each generation, and (iii) it adds an embedding‑reuse surrogate together with a first‑principles oracle‑fidelity study. A separate and more established line steers crystal generators by reinforcement‑learning fine‑tuning of the model weights --- MatInvent \citep{chen2025accelerating}, CrystalFormer‑RL \citep{cao2026reinforcement}, and related policy‑gradient methods \citep{park2026guiding, hollmer2026open} --- and, concurrently, a Gaussian‑process acquisition gate has been inserted into such an RL loop using pretrained embeddings \citep{alvi2026surrogate}. DiffGA is deliberately orthogonal to this line. It applies no gradient updates, reinforcement learning, or fine‑tuning of any kind, steering a frozen model entirely at inference, and its surrogate reuses the force field's own embedding rather than gating a separately trained policy.

Several limitations bound the scope of these results, and addressing them constitutes the bulk of ongoing and future work. Most fundamentally, DiffGA is only as good as its oracle. The absolute quality of any hit is capped by the fidelity of the oracle, regardless of whether is uses machine‑learned force field or DFT. Our validation quantifies this rather than obscuring it. Second, our first‑principles validation is a representative sample rather than an exhaustive re‑evaluation, and we deliberately do not propose or fully characterize specific new compounds. The validated carbides (e.g., WC, CrC) are polymorphs within a known chemical space of ultra‑hard materials and, thus, confirm that the method reaches the genuinely stiff region of chemical space. Detailed analysis of structural or polymorphic novelty is left to dedicated follow‑up. Third, the genetic operators employed here --- cut‑and‑splice crossover and electronegativity‑guided substitutions (see Genetic operators in the Methods section) --- are deliberately simple, and more expressive, symmetry‑aware operators are a clear avenue for improvement. Fourth, while the additional cost associated with repeated denoising is negligible compared to the cost of the oracle (see Computational cost under Methods), it could be replaced by Tweedie's formula \citep{efron2011tweedies, kim2021noise2score} to further accelerate the framework. Fifth, the on‑the‑fly surrogate presupposes a force field that exposes a usable structural embedding. While this is a property shared by essentially all modern MLFFs, it is not guaranteed in general. Finally, all results here are computational. Experimental synthesis and characterization, together with coupling DiffGA to end‑use feasibility oracles such as synthesizability or electrochemical accessibility, are important next steps. Crucially, most of these limitations concern the oracle and the downstream validation burden rather than the DiffGA mechanism itself, which is agnostic to both the oracle and the scorer, and admits drop‑in replacements by more accurate or higher‑fidelity oracles as they become available.

\section*{Methods}
\label{sec:methods}

\subsection*{Problem setup and the DiffGA framework}
\label{subsec:problem_setup_and_diffga}
We build on MatterGen \cite{zeni2025generative} as the diffusion backbone $\mathcal{D}$, which produces candidate crystal structures by iterative denoising from $t = 1$ (pure noise) to $t = 0$ (candidate). DiffGA modifies MatterGen only at inference, by interleaving evolutionary operators over partially denoised states. We define:
\begin{enumerate}
    \item \textbf{Horizon time} $t_h \in (0,1)$: an intermediate timestep at which denoising is paused. The partially denoised structures are referred to as the set of horizon structures, $\mathbf{X}^{(h)}$.
    \item \textbf{Number of generations} $G$: the number of evolutionary iterations applied to the horizon population.
    \item \textbf{Population multiplier} $P$: the factor by which the horizon population is scaled.
    \item \textbf{Final structures} $\mathbf{X}^{(f)}$: fully denoised structures obtained by continuing diffusion from horizon states.
    \item \textbf{Oracle} $S$ : $\mathbf{X}^{(f)} \mapsto \mathbb{R}$: any user-defined metric which could be black-box, differentiable, non-differentiable, analytical, or a combination thereof.
    \item \textbf{Surrogate} $\hat{S}$: $\mathbf{X}^{(f)} \mapsto \mathbb{R}$: a cheap GP model trained on-the-fly during a DiffGA run to mimic the oracle, yielding predictive mean and uncertainty.
    \item A scoring function which maps oracle (or surrogate) evaluated properties to $[0,1]$.
\end{enumerate}

Let $\mathcal{P}_g = \{\mathbf{X}^{(h)}_i\}_{i=1}^{B}$ be the population of $B$ horizon structures at generation $g$. The procedure is:
\begin{enumerate}
    \item \textbf{Partial Denoising}: The initial horizon population $\mathcal{P}_{g=1}$ is produced by the native diffusion model, $\mathcal{D}$, by denoising every structure up to the horizon time $t_h$.
    \begin{align*}
        \mathbf{X}_i^{(h)}=\mathcal{D}_{1\rightarrow t_h}(\mathbf{z}_i), \quad \mathbf{z}_i\sim\mathcal{N}(0,I)
    \end{align*}
    where $\mathcal{D}_{1\rightarrow t_h}$ denotes the iterative diffusion sampling process run for all timesteps from $t=1$ to $t=t_h$. Subsequent horizon populations $\mathcal{P}_g\;\forall\;g\in\{2,...,G\}$ are obtained via evolutionary operations and elitist retention, as described in step 6.
    \item \textbf{Population multiplier}: For the first generation the horizon population is replicated $P$ times. A larger population, even if less diverse, is often expected to perform better in GA frameworks.
    \item \textbf{Leftover Denoising and Scoring}: Each horizon structure is fully denoised to produce a physical structure
    \begin{align*}
        \mathbf{X}_i^{(f)}=\mathcal{D}_{t_h\rightarrow0}\left(\mathbf{X}_i^{(h)}\right)
    \end{align*}
    which then undergoes structural relaxation. Note that for the purpose of target optimization, properties are first computed by the oracle, the results of which are used by the scorer to obtain a score between $0$ and $1$.
    \begin{align*}
        s_i=\text{Scorer}\left(S\left(\mathbf{X}_i^{(f)}\right)\right)
    \end{align*}
    This score is associated back to the horizon structure to form the set of triplets $z_g = \{(\mathbf{X}^{(h)}_i, \mathbf{X}^{(f)}_i, s_i)\}_{i=1}^{PB}$. If duplicate final structures are detected (checked via \texttt{pymatgen}'s \texttt{StructureMatcher} with default parameters), the corresponding horizon structures are mutated and re-denoised until unique or a retry limit is reached.
    \item \textbf{Surrogate training set}: If the surrogate-accelerated pipeline is being used, the pair $\{\mathbf{X}_i^{(f)}, S\left(\mathbf{X}_i^{(f)}\right)\}$ is added to the training set for the surrogate, $\hat{S}$.
    \item \textbf{Surrogate readiness}: If $\hat{S}$ is ready to replace the oracle after a rank-correlation gate, (i) it replaces the oracle in step $3$, (ii) scoring in step $3$ is done under an uncertainty-aware acquisition function, and (iii) the set of points to be added to the training set in step $4$ are reduced to a fraction $\rho$ of its most uncertain predictions, re-evaluated by $S$. Details can be found under the section on the surrogate in Methods.
    \item \textbf{Selection with elitist retention}: Horizon structures are ranked by score, but elites from the previous generation are included to guarantee monotonic retention of the best solutions.
    \begin{align*}
        z^\text{elite}_g = \mathrm{TopK}(z_g \cup z^\text{elite}_{g-1})
    \end{align*}
    \item \textbf{Evolutionary operations}: Genetic operators (mutation, crossover) are applied to the $k$ retained horizon structures to produce the next population of $P \times B$ horizon structures, injecting a directional, property-driven bias into the denoising process.
    \begin{align*}
        \mathcal{P}^\text{elite}_g &= \{\mathbf{X}^{(h)} | \mathbf{X}^{(h)} \in z^\text{elite}_g \} \\ 
        \mathcal{P}_{g+1} &= \mathrm{GA\_Update}(\mathcal{P}^\text{elite}_g)
    \end{align*}
    \item \textbf{Surrogate fitting}: $\hat{S}$ is refit on the entire set of its expanded training set for that generation.
    \item Steps 3–8 are repeated for $G$ generations or until an early stopping criterion is reached. At the final generation, all horizon structures are fully denoised to yield the final candidates.
\end{enumerate}

The pseudocode for the entire framework can be found in Algorithm~\ref{alg:diffga}. The limit $t_h \to 0$ is operationally equivalent to a GA applied to fully denoised structures, whereas the generation $g = 1$ is identically distributed to base-model sampling (section S1 in the SI).

\begin{algorithm}[!t]
   \caption{The DiffGA workflow}
   \label{alg:diffga}
\begin{algorithmic}
   \STATE {\bfseries Input:} Initial noisy population at the start of a diffusion denoising trajectory, $\{\mathbf{X}_i^{(0)}\}_{i=1}^B$
   \STATE \underline{Prerequisites}: Oracle, $S$, and a scoring function
   \STATE Partially denoise $\{\mathbf{X}_i^{(0)}\}_{i=1}^B$ $\longrightarrow$ $\{\mathbf{X}_i^{(h)}\}_{i=1}^B$
   \STATE Set $\mathcal{P}_1$ as $\{\mathbf{X}_i^{(h)}\}_{i=1}^{B}$ replicated $P$ times
   \FOR{$g=1$ {\bfseries to} $G$}
   \STATE Initialize $z$ as an empty list
   \STATE Complete denoising for $\mathcal{P}_g \rightarrow \{\mathbf{X}_i^{(f)}\}_{i=1}^{PB}$
   \FOR{$j=0$ {\bfseries to} $PB-1$}
   \IF{$\mathbf{X}_j^{(f)}$ is \textit{not} unique in $z$}
   \STATE Mutate $\mathbf{X}_j^{(h)}$ until $\mathbf{X}_j^{(f)}$ is unique in $z$ or a maximum number of trials is reached
   \ENDIF
   \IF{$\hat{S}$ is ready}
   \STATE $\hat{p}_j, \hat{\sigma}_j=\hat{S}\left(\mathbf{X}_j^{(f)}\right)$
   \STATE $s_j=$ Highest score attainable within $[\hat{p}_j-\beta\hat{\sigma}_j,\hat{p}_j+\beta\hat{\sigma}_j]$
   \ELSE
   \STATE $s_j=\text{Scorer}\left(S\left(\mathbf{X}_j^{(f)}\right)\right)$
   \ENDIF
   \STATE Append the triplet $\left(\mathbf{X}_j^{(h)}, \mathbf{X}_j^{(f)}, s_j\right)$ to $z$
   \ENDFOR
   \IF{$\hat{S}$ is ready}
   \STATE Query oracle $S$ for fraction $\rho$ of most uncertain $\hat{S}$ predictions to get $\{\mathbf{X}_k^{(f)}, S(\mathbf{X}_k^{(f)})\}_{k=1}^{\rho PB}$
   \STATE Add $\{\mathbf{X}_k^{(f)}, S(\mathbf{X}_k^{(f)})\}_{k=1}^{\rho PB}$ to $\hat{S}$'s training set
   \ELSE
   \STATE Add $\{\mathbf{X}_k^{(f)}, S(\mathbf{X}_k^{(f)})\}_{k=1}^{PB}$ to $\hat{S}$'s training set
   \ENDIF
   \STATE Extract top-$k$ candidates from $z$ with elitist retention as $z^\text{elite}_g = \mathrm{TopK}(z_g \cup z^\text{elite}_{g-1})$
   \STATE Set $\mathcal{P}^\text{elite}_g = \{\mathbf{X}^{(h)} | \mathbf{X}^{(h)} \in z^\text{elite}_g \}$
   \IF{$g<G$}
   \IF{early stopping criteria is met}
   \STATE Return $\mathcal{P}_g$
   \ELSE
   \STATE $\mathcal{P}_{g+1}=\text{GA\_Update}\left(\mathcal{P}_g^{\text{elite}}\right)$
   \STATE Refit $\hat{S}$ on all training data
   \ENDIF
   \ELSE
   \STATE Return $\mathcal{P}_G$
   \ENDIF
   \ENDFOR
   \STATE {\bfseries Output:} Target-specific, high-scoring set of candidates, $\{\mathbf{X}_i^{(f)}\}_{i=1}^{B}$, and a surrogate, $\hat{S}$
\end{algorithmic}
\end{algorithm}

\subsection*{Genetic operators}
\label{subsec:genetic_ops}
\begin{figure*}[htb]
\centering
\includegraphics[width=\textwidth]{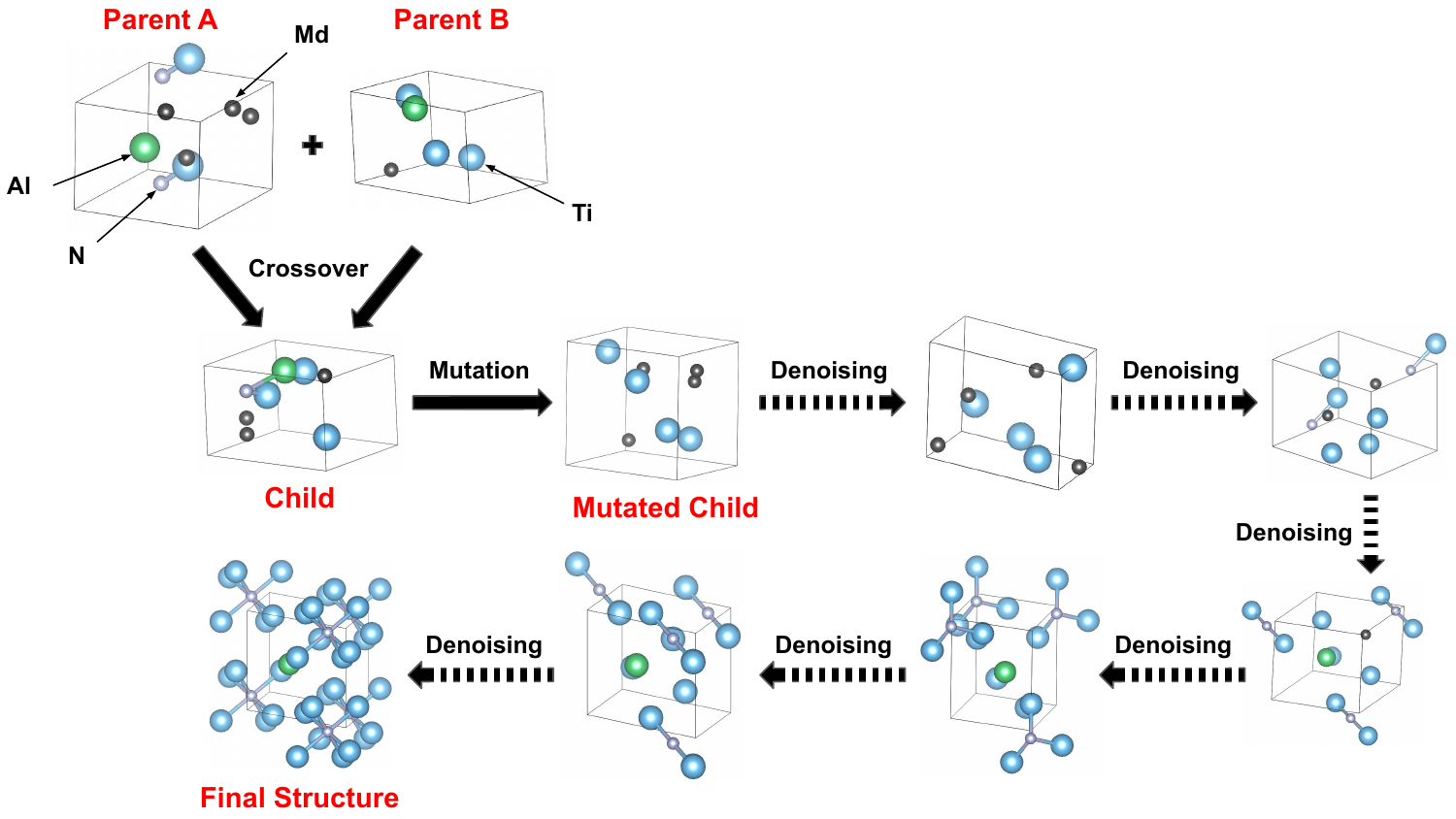}
\caption{Visual representation of how a single structure is obtained via evolutionary operations and subsequent denoising. Here, $t_h=0.5$. Note that `Parent A', `Parent B', `Child', and `Mutated Child' all exist at $t=t_h$. The \texttt{Md} atoms are noisy placeholders introduced by the underlying diffusion model for atom types yet to be fully denoised.}
\label{fig:ga_ops}
\end{figure*}

We use a cut-and-splice crossover inspired by USPEX \cite{glass2006uspex} and ASE \cite{hjorth2017atomic}, and mutation operators adapted from ASE with modifications to enforce a directional bias under chemical-space constraints, including species substitutions guided by electronegativity similarity. Thus, through a combination of these operations, favorable atom types can be artificially introduced into the atomic structure which, through denoising across generations, can result in physically valid and meaningful structures. While these simple operators define the GA component in this work, more expressive, symmetry-aware operators are left to future work. Fig.~\ref{fig:ga_ops} showcases an example of how the aforementioned evolutionary operations modify a given structure, including snapshots from the subsequent denoising trajectory, using an example from experiment A.

\subsection*{Scoring function}
\label{subsec:scorer}
Scoring functions act upon oracle evaluations of fully denoised structures and return a scalar between $0$ and $1$. Let $p_t$ be the target property, $p_i$ the computed property, and $\sigma$ the allowed tolerance. We use:
\begin{enumerate}
    \item Gaussian:
    \begin{align*}
            s_i=S(\mathbf{X}^{(f)}_i)=e^{\frac{-\left(p_i-p_t\right)^2}{2\sigma^2}}
    \end{align*}
    \item Huber-based \cite{huber1992robust}:
    \begin{align*}
            s_i&=S(\mathbf{X}^{(f)}_i)=\frac{1}{1+l} \\
            l&=\begin{cases}
                0.5\times(p_i-p_t)^2 & \lvert p_i-p_t \rvert \leq \sigma \\
                \sigma\left(\lvert p_i-p_t \rvert - \frac{\sigma}{2}\right) & \lvert p_i-p_t \rvert > \sigma
            \end{cases}
    \end{align*}
\end{enumerate}

Specifically, experiment A uses the Gaussian score on shear modulus (the \texttt{chemical\_system} model guarantees the chemical constraint). Experiment B uses a composite score, $s_i = 0.5 s_i^{(1)} + 0.5 s_i^{(2)}$, where $s_i^{(1)}$ is the Gaussian property score and $s_i^{(2)}$ is a chemical-validity score rewarding overlap with the target chemical space and penalizing out-of-space elements. Experiment C mimics this but replaces the Gaussian with the Huber-based term which has a smoother decay and is more suited to maximization with a high nominal target, along with a chemical-validity score enforcing a two-element, single-transition-metal carbide.

\subsection*{Oracle}
\label{subsec:oracle}
The typical oracle setup in this study is a structural relaxation step followed by property evaluation. Structural relaxations and single-point energies use MatterSim, with a force threshold of $0.05$ eV/atom for the former. The energy above the hull is computed with the same procedure and reference framework as MatterGen \cite{zeni2025generative} for consistency. Elastic tensors are evaluated through MatTen \cite{wen2024equivariant}. At this point, it is important to note that DiffGA is agnostic to these choices. Any relaxation method, force field, or property predictor (including DFT, analytical heuristics, or experiment) can be substituted, with the choice only affecting numerical accuracy but not the framework. The goal of this study is not to benchmark MLFFs or to characterize novel materials, but rather to demonstrate DiffGA independent of the evaluation mechanism.

\subsection*{On-the-fly surrogate and uncertainty-aware acquisition}
\label{subsec:surrogate}
\begin{figure*}[htbp]
\centering
\subfloat[Experiment A]{\includegraphics[width=0.31\linewidth]{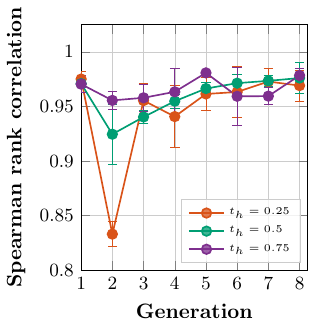}\label{subfig:relaxed_ion_dft}}
\hfill
\subfloat[Experiment B]{\includegraphics[width=0.31\linewidth]{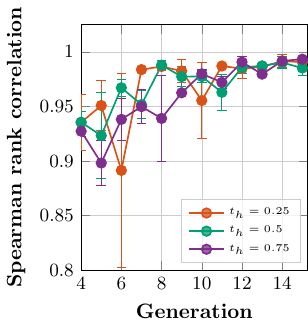}\label{subfig:clamped_ion_dft}}
\hfill
\subfloat[Experiment C]{\includegraphics[width=0.31\linewidth]{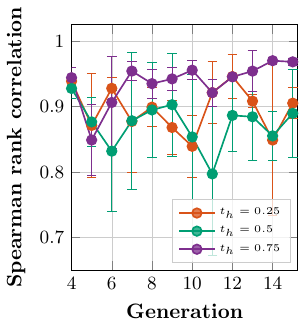}\label{subfig:ehull_dft}}
\caption{Spearman rank correlation between surrogate-predicted and true oracle properties evaluated on the most recent generation, for all experiments and $t_h$ values. Rank correlations remain high throughout the campaigns, demonstrating that the surrogate preserves the candidate ranking required by the evolutionary selection step. Markers show the mean across all seeds and error bars denote the standard deviation across seeds.}
\label{fig:bo_spearman_rank_corr}
\end{figure*}

To reduce oracle cost during a run, DiffGA optionally trains a surrogate $\hat{S}$ on the \{structure, oracle-evaluated property\} pairs it accumulates, and delegates oracle evaluations to $\hat{S}$ once it is trustworthy. This is expected to approximately halve the total expensive property‑evaluation costs per discovery campaign (see Fig.~\ref{fig:bo_performance}), with the additional benefit of having a well-trained surrogate to use or fine-tune externally, if needed.
\begin{enumerate}
    \item Features: Every candidate is relaxed by an MLFF (here, MatterSim) before being passed on to the oracle. MatterSim is an M3GNet-class model \cite{chen2022universal} from which we extract its internal graph representation, i.e., the per-atom features entering the energy head, mean-pooled over atoms, as a fixed $d$-dimensional descriptor ($d = 256$ for the \texttt{mattersim-5M} model used here). The descriptor is obtained from the already-relaxed structure with a single additional forward pass (negligible next to the relaxation itself) and requires no gradients or training of the force field. Because the same MLFF supplies both the relaxation and the descriptor, the surrogate introduces no new learned representation to maintain. Note that this technique is independent of the choice of MLFF.
    \item Model: $\hat{S}$ is an exact Gaussian process (GP) with a constant mean and a scaled Mat\'{e}rn-5/2 kernel with automatic relevance determination (one length-scale per input dimension) over the standardized descriptor. The target (oracle property) is normalized to zero mean and unit variance. The GP yields a predictive posterior with mean $\mu(\mathbf{X})$ and standard deviation $\sigma(\mathbf{X})$. Only the GP hyperparameters (length-scales, output scale, mean, observation noise) are optimized by maximizing the exact marginal likelihood. It is important to note that the surrogate predicts the physical property, not the score. The cheap, known scoring function is applied exactly to the predicted property, keeping the learned target low-dimensional and physically meaningful and letting one surrogate serve any score defined on that property.
    \item Prospective rank-correlation gate: After each refit, we assess the surrogate not by fit quality on seen data but by its ability to rank the next batch of candidate structures. We hold out the most-recent generation's points as a validation block and compute the GP's leave-block-out predictive means in closed form (the block generalization of the leave-one-out identity), then score them by Spearman rank correlation against the true oracle. We use rank correlation rather than $R^2$ because (i) the GA consumes only the relative order of candidates, and (ii) as the search concentrates, the property variance within a generation shrinks, which mechanically collapses $R^2$ even when ranking remains accurate --- a failure mode we observed directly. The surrogate is promoted to replace the property oracle only once it has accumulated at least $n_\text{ready}$ points and its held-out rank correlation exceeds $\tau_\text{rank}$ (defaults used in the current study were $n_\text{ready}=150$, $\tau_\text{rank}=0.5$). If the surrogate fails to satisfy these conditions, the true oracle continues to evaluate structures and the new point is added to the training set. A refit that turns anti-predictive (negative rank correlation) is rolled back to the last surrogate that passed the gate, so a bad refit can never degrade the deployed model. Fig.~\ref{fig:bo_spearman_rank_corr} shows that the Spearman rank correlation for the surrogate across generations for each experiment and value of $t_h$, averaged across all seeds, remains high, proving the validity of the surrogate.
    \item Acquisition function: When the surrogate evaluates a candidate, we do not select on the predicted property alone. Instead, we apply an upper-confidence-bound-style optimism \cite{shahriari2016taking, snoek2012practical} that is agnostic to the objective's shape. For predicted property $\hat{p}$ with uncertainty $\hat{\sigma}$, we consider the band $[\hat{p} - \beta\hat{\sigma},\, \hat{p} + \beta\hat{\sigma}]$ and assign the candidate the property corresponding to the best score attainable within it (evaluated at the band edges for target maximization or minimization and, for target-seeking objectives, at the in-band point nearest the target). This reduces exactly to the mean predicted property when $\hat{\sigma}=0$, i.e., identical to standard DiffGA. Moreover, this correctly rewards uncertain candidates for both maximization objectives (Experiment C) and peaked target-seeking objectives (Experiments A, B) with no objective-specific tuning. We use $\beta = 2$.
    \item Active learning and drift control: Uncertainty also directs where real oracle calls are spent. Each generation, a fraction $\rho$ of surrogate-evaluated candidates with the highest predictive uncertainty is re-evaluated with the true oracle and added to the training set (default $\rho = 0.3$). The surrogate is then refit once per generation on all accumulated data. This continually supplies the labels the model is least sure about and re-anchors it to the true oracle as the population moves, guarding against distributional drift and surrogate blind spots.
\end{enumerate}

To measure fidelity retention cleanly, we run matched pairs from an identical first-generation horizon population: one track guided entirely by the true oracle, one by the surrogate-accelerated pipeline, with the random-number state restored to the same value before each track. The two runs therefore diverge only through the selection signal (true property vs surrogate-predicted property under optimistic acquisition), isolating the surrogate's effect on the search from run-to-run sampling noise. The DiffGA loop, objective, and elitist selection are otherwise unchanged. The surrogate affects only whether the expensive property step is computed or predicted, and how predictive uncertainty enters selection.

\subsection*{DFT details}
\label{subsec:dft}
We validate a representative sample of structures generated in this study with the projector‑augmented‑wave method \cite{mortensen2005real} in VASP \cite{kresse1996efficient}. Two protocols are used, matched to their purpose. For oracle‑fidelity comparisons against the (PBE‑level) MLFF --- shear modulus and Cs–Sn–I energy above the hull --- candidates are relaxed with MP‑compatible GGA settings. We use \texttt{pymatgen}'s \texttt{MPRelaxSet} with PBE, an energy cutoff of $520$ eV, $\Gamma$‑centered k‑meshes consistent with the mesh density reported in MP, and run spin‑polarized calculations, wherever appropriate. To compute the energy above the hull for Cs-Sn-I structures, they are then placed on a GGA/GGA+U hull built from MP reference entries under a single, functional‑consistent correction scheme \citep{wang2021framework}. For the absolute stability of the carbide hits, where no MLFF comparison is made, candidates are instead relaxed with r2SCAN level of theory with an energy cutoff of $680$ eV, and placed on a pure r2SCAN convex hull of the corresponding chemical system, which is self‑consistent for novel compositions.

Specifically, for elastic tensor calculations, a full cell relaxation with a force threshold of $5\times10^{-3}$ eV/\AA) is run, and the elastic tensors are then computed by finite differences. VASP reports both the rigid‑ion (clamped) tensor and the relaxed-ion tensor which are used to compute the Voigt–Reuss–Hill \citep{hill1952elastic} shear modulus. By construction the relaxed‑ion modulus is bounded above by the clamped‑ion modulus.

\subsection*{Computational cost}
\label{subcec:comp_cost}
The computational cost in DiffGA is dominated by the oracle. Each structure is initially denoised from random noise down to $t_h$, after which it participates in the genetic loop. Every structure is evaluated once per generation, after completing its remaining denoising steps. Let $C_{\text{oracle}}$ denote the cost of a single oracle call. Then, the total number of oracle calls is:
\begin{align*}
    N_{\text{oracle}}=P\times B\times G
\end{align*}
and, thus, the total cost associated with the oracle is:
\begin{align*}
    C_{\text{oracle}}^{\text{total}}=P\times B\times G\times C_{\text{oracle}}
\end{align*}
This, so far, is identical to a standard GA pipeline operating on the same population and number of generations, and independent of the diffusion model used.

The diffusion-related costs are almost entirely dependent on the horizon time, $t_h$, and the number of generation, $G$. That is because the larger these parameters, the larger the amount and number of times the leftover denoising needs to be performed for every structure. Let $C_{\text{diff}}$ denote the total cost of completely denoising a single structure. Then the total diffusion cost can be approximated as:
\begin{align*}
    C_{\text{diff}}^{\text{DiffGA}}&\approx P\times B\times C_{\text{diff}}\times[(1-t_h)+Gt_h]\\&=P\times B\times C_{\text{diff}}[1+t_h(G-1)]
\end{align*}
This shows that the diffusion cost scales linearly with respect to the population size, horizon time, and number of generations, and remains predictable and bounded. Moreover, the only additional cost for diffusion comes from the second term, which is proportional to $Gt_h\;\forall\;G>1$.

To further reduce the computational cost of DiffGA, we introduce an optional control that delays expensive oracle evaluations until a specified generation. This is particularly effective in settings where the scoring function is composite and includes costly property evaluations along with a check on chemical space validity, as in Experiments B and C. Concretely, for the first $n$ generations, DiffGA can operate using a partial oracle and scoring function that evaluate only inexpensive structural criteria (e.g., chemical system validity or compositional constraints), while deferring expensive property-based oracle calls and associated scoring to the remaining $G-n$ generations. For example, when using the unconditioned \texttt{mattergen\_base} model, early generations can be dedicated solely to steering the population toward the desired chemical space, after which property evaluations (e.g., energy above hull or elastic moduli) are introduced once the chemical space search has been sufficiently narrowed.

The on‑the‑fly surrogate further reduces the property-evaluation component of $C_\text{oracle}$. Once the surrogate is promoted (after $n_\text{ready}$ points and a passing prospective rank‑correlation gate), the expensive property call is issued only for the fraction $\rho$ of most‑uncertain candidates re‑queried for active learning each generation. The number of expensive property evaluations therefore scales as
$$N_\text{prop}^\text{SA} \approx PB\times[1 + \rho(G-1)]$$
versus $PBG$ for the full‑oracle pipeline, a reduction that grows with $G$ and, for $\rho = 0.3$, exceeds a factor of three at the generation counts used here (see Fig.~\ref{fig:bo_performance}). This is complementary to the deferred‑oracle control above which avoids expensive calls in early generations.

\section*{Contributions}
S.A. performed idea conceptualization, framework development, experimental design, ablations, data analysis and visualization, hypothesis testing, writing, and editing. K.R. helped in idea conceptualization and experimental design. R.W. and A.R.S. provided project leadership and helped with idea conceptualization. R.W. also helped with code validation. All authors contributed towards manuscript editing, review and have read and approved the manuscript.

\section*{Acknowledgment}
The authors acknowledge SandboxAQ leadership, especially Ang Xiao, Adam Lewis, Takeshi Yamazaki, Nadia Harhen, and Jack Hidary, for their support of this research. The authors thank Tyler Sours for valuable discussions and feedback. The authors recognize computational and engineering support from SungYeon Kim, Daniel Critchfield, and J.P. Murgueitio. This research received no specific grant from any funding agency.



\bibliography{main}

\newpage
\section*{Supplementary Information}

\section{Genetic algorithm and diffusion baselines}
\label{sec:ga_diff_baselines}

\subsection{Genetic Algorithm}
In this section, we justify the use of a small horizon time, $t_h=0.002$, as a proxy for a genetic algorithm (GA) baseline in our experiments.

Consider the DiffGA procedure operating with a diffusion model $\mathcal{D}$, a horizon time $t_h$, and a GA loop applied to horizon structures. As the horizon time approaches zero, i.e., $t_h\rightarrow0$, the DiffGA procedure becomes operationally equivalent to a genetic algorithm applied directly to fully denoised structures, up to a negligible final denoising step. This is because DiffGA operates by first denoising samples up to a specified horizon time $t_h$, producing a population of horizon structures. These horizon structures then enter a GA loop, where for each generation:
\begin{enumerate}
    \item Each horizon structure is fully denoised to produce a physical structure.
    \item The resulting structure is evaluated using a scoring function.
    \item The score is assigned back to the corresponding horizon structure.
    \item Selection, mutation, and crossover are applied at the horizon level to produce the next generation of horizon structures.
\end{enumerate}
When $t_h$ is sufficiently small, nearly the entire diffusion trajectory is completed \textit{before} the GA loop begins. In this regime, the remaining denoising applied during each GA generation consists of only a single or handful of diffusion steps. Specifically, for this study, a horizon time of $t_h=0.002$ corresponds to a single diffusion step during each GA generation. As a result:
\begin{enumerate}
    \item The diffusion model no longer meaningfully shapes the population during the evolutionary process.
    \item The dominant dynamics are governed by selection and genetic variation.
    \item The diffusion model acts only as an initial population generator for the first generation.
\end{enumerate}
Thus, for practical purposes, DiffGA with a very small horizon time behaves like a GA workflow operating on generated structures, with the diffusion component contributing negligible additional structural guidance. We therefore report results at $t_h=0.002$ as a GA baseline, enabling a controlled comparison between pure evolutionary search and diffusion-guided evolutionary search (DiffGA). This choice allows us to isolate the contribution of horizon-guided diffusion to both hit rates and structural diversity.

\subsection{Diffusion}
We also formalize the equivalence relationship between the first generation ($g=1$) of DiffGA and standard diffusion sampling, and provide further details about the compute-matched oracle baseline for diffusion. In the DiffGA framework, genetic operations are performed at the horizon level to create the horizon population for subsequent generations. Crucially, however, at the first generation, no genetic operations have yet been applied. The horizon structures at this stage are sampled directly from the diffusion prior, and each is fully denoised using the same diffusion dynamics as in standard sampling. Since neither selection nor variation has occurred, the outputs from the first generation depend solely on the diffusion process.
    
As a result, outputs from $g=1$ are equivalent to samples drawn from the base diffusion model, subject only to the same conditioning (if any) used during diffusion sampling. This equivalence holds for all values of the horizon time as the full denoising trajectory is unchanged prior to any evolutionary feedback, and establishes generation $g=1$ as a true pure-diffusion baseline ($\text{Diff.}^\dagger$).

Furthermore, sampling $N=PBG$ structures from the pretrained diffusion model and ranking them by the scorer $S$ constitutes a compute-matched diffusion baseline ($\text{Diff.}^\ddagger$) for DiffGA and isolates the effect of oversampling. In DiffGA, each generation produces a population of intermediate states that are fully denoised and evaluated using a scoring function $S$. Across $G$ generations, and assuming a population multiplier of $P$, this results in a total of $N=PBG$ oracle calls and sampled structures, where $B$ is the batch size. These evaluations are used iteratively to guide evolutionary selection and modification of intermediate representations.

In the compute-matched diffusion baseline considered in this study (see Table 1 in the main manuscript), we instead draw $N$ independent samples from the same pretrained diffusion model without evolutionary feedback. Each sample is already fully denoised and directly evaluated using the same oracle $S$, and the top-scoring candidates are retained. By construction, both procedures incur the same number of oracle evaluations and rely on the same generative prior and scoring mechanism. The key difference is that this diffusion baseline applies oracle scores \textit{post hoc} to rank independently sampled candidates, whereas DiffGA uses oracle feedback iteratively to bias subsequent sampling through evolutionary operations. This baseline tests whether improvements observed with DiffGA can be attributed solely to increased sampling and oracle access, or whether they arise from the structured use of oracle feedback during generation. Empirically, we find that DiffGA consistently outperforms this compute-matched diffusion baseline, indicating that iterative evolutionary feedback drives the observed gains.

\newpage
\section{Score and property evolution across generations}
\label{sec:scores_and_props}
This section presents an exhaustive set of plots depicting raw DiffGA performance across all generations and horizon parameters for every experiment. In each of Figs.~\ref{fig:all_TiAlN_shear_boxplots}, \ref{fig:all_CsSnI_chemsys_e_boxplots}, \ref{fig:all_CsSnI_chemsys_boxplots}, \ref{fig:all_CsSnI_base_boxplots}, and \ref{fig:all_tmcarbides_boxplots}, every sub-figure contains results from a unique value of $t_h$. Moreover, within every sub-figure, there exist two plots:
\begin{itemize}
    \item On the left, we show the distribution of DiffGA scores across generations for the three population multipliers ($P$) considered in this study. Along with the full distributions, we report both the mean and the maximum score at each generation to provide additional context. We also highlight, using a lighter shade, the top 25\% of scores. This choice reflects the fact that conclusions drawn from the entire distribution can be misleading in realistic materials discovery settings, where success is often driven by the presence of a small subset of high-quality candidates that satisfy the target constraints, rather than by average performance. Note that for results pertaining to the use of the \texttt{mattergen\_base} model, we also plot the percentage of structure satisfying the chemical space constraint, labeled as `\% ChemSys'.
    \item On the right, we show a similar distribution of corresponding property values across generations. As with the scores, we report the mean and the best property at each generation, as well as a separate distribution representing the properties corresponding to the top $25\%$ of scores. Additionally, we also include the target property (wherever applicable) in these plots with a dashed line. For plots pertaining to experiment C, we apply an upper limit of 300 GPa and, for purposes of this study, assume that any structures with a predicted shear modulus greater than this limit are erroneous or invalid.
\end{itemize}

Across all experiments, we observe a consistent increase in mean DiffGA scores over successive generations, indicating steady improvement of the population under evolutionary refinement. Importantly, for hit-finding tasks -- where the objective is to identify \textit{at least} one structure satisfying the target constraints -- only a small number of generations is sufficient to obtain high-scoring candidates. This highlights DiffGA’s ability to rapidly steer the generative process toward favorable regions of the design space, even before full population-level convergence is achieved.

In some cases, distributions for certain generations appear missing or show sudden drops in the mean score. This is an artifact of early stopping. When a pre-specified convergence criterion is met (such as when $80\%$ of the total structures generated in a given generation have scores greater than $0.85$), the evolutionary algorithm halts for that seed. Because results are averaged across three seeds, if the higher-scoring seeds terminate early, their scores no longer contribute to the mean for later generations. The remaining seeds, which continue, may have lower scores, producing the apparent dip. Examination of individual seed trajectories confirms that this behavior is consistent with early stopping and does not indicate a failure of DiffGA. The above argument can be further strengthened by observing that the top $25\%$ of the scores always increase across generations, irrespective of early stopping. In places where this quirk of the plots is more apparent (such as in Figs.~\ref{fig:CsSnI_chemsys_e_0p002}, \ref{fig:CsSnI_chemsys_0p98}, \ref{fig:CsSnI_base_0p002}), we include a plot for individual performances across the different seeds for clarity (such as, Figs.~\ref{fig:CsSnI_chemsys_e_0p002_seeds}, \ref{fig:CsSnI_chemsys_0p98_seeds}, \ref{fig:CsSnI_base_0p002_seeds}).

\newpage

\begin{figure}[ht!]
\centering
\subfloat[$t_h=0.002$]{
\includegraphics[width=0.95\textwidth]{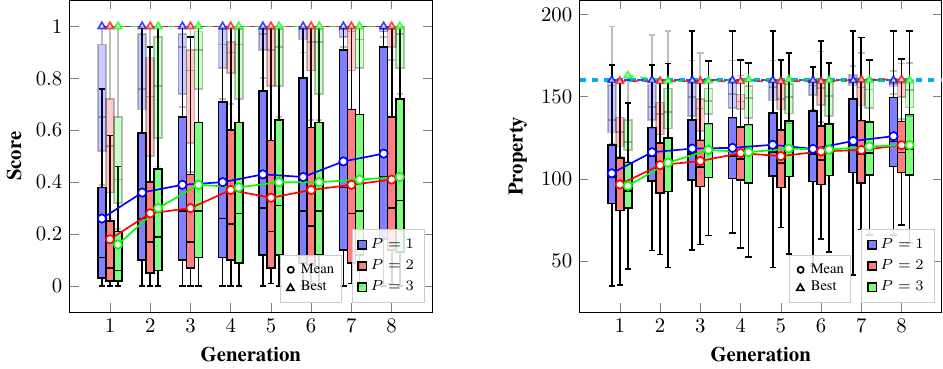}
\label{fig:TiAlN_shear_0p002}
}
\vspace{0.35cm}
\subfloat[$t_h=0.25$]{
\includegraphics[width=0.95\textwidth]{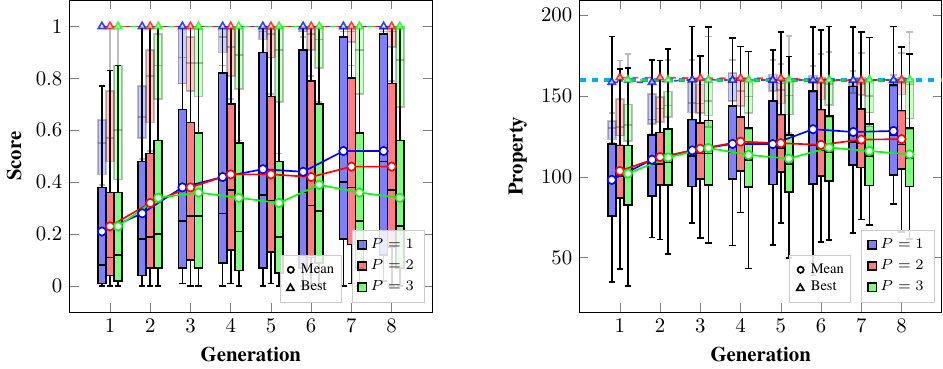}
\label{fig:TiAlN_shear_0p25}
}
\vspace{0.35cm}
\subfloat[$t_h=0.5$]{
\includegraphics[width=0.95\textwidth]{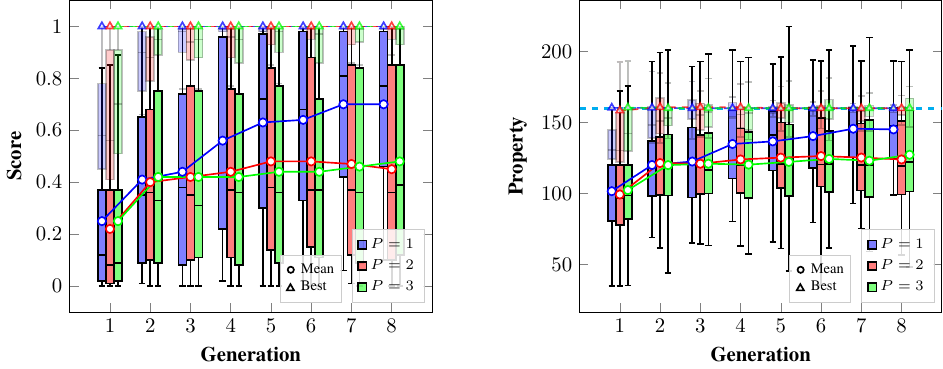}
\label{fig:TiAlN_shear_0p5}
}
\end{figure}

\begin{figure}[ht!]\ContinuedFloat
\centering
\subfloat[$t_h=0.75$]{
\includegraphics[width=0.95\textwidth]{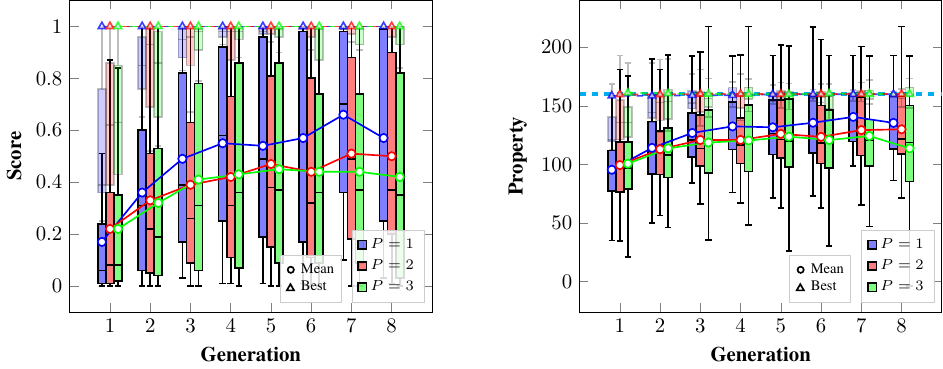}
\label{fig:TiAlN_shear_0p75}
}
\vspace{0.35cm}
\subfloat[$t_h=0.98$]{
\includegraphics[width=0.95\textwidth]{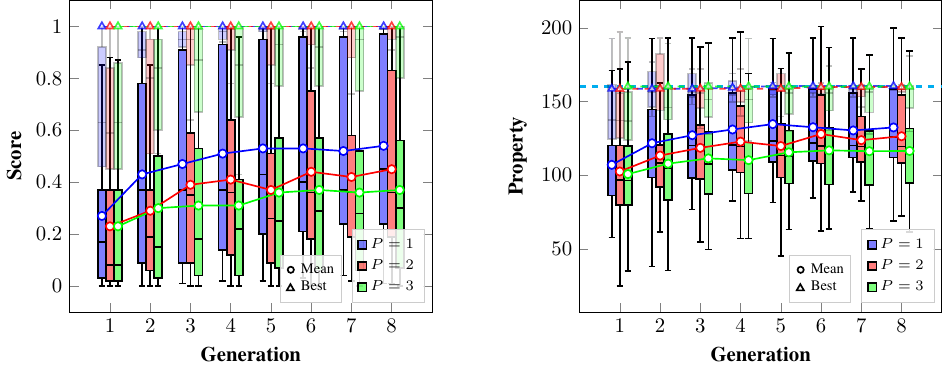}
\label{fig:TiAlN_shear_0p98}
}
\caption{Distribution of scores \& properties across generations for experiment A (\texttt{chemical\_system} model).}
\label{fig:all_TiAlN_shear_boxplots}
\end{figure}

\newpage

\begin{figure}[ht!]
\centering
\subfloat[$t_h=0.002$]{
\includegraphics[width=0.95\textwidth]{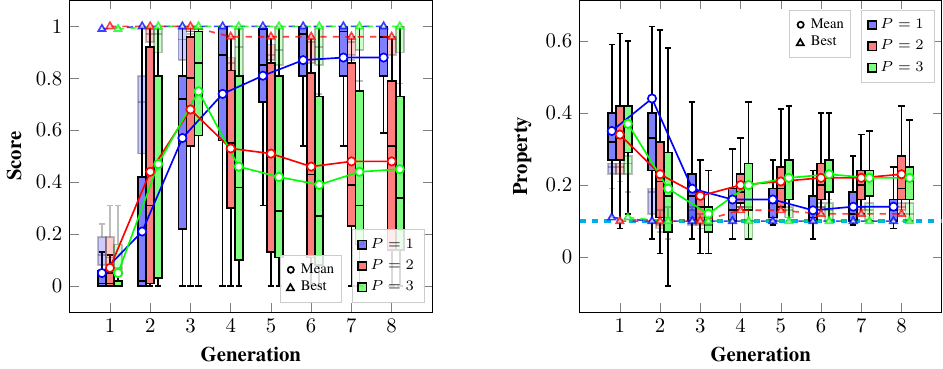}
\label{fig:CsSnI_chemsys_e_0p002}
}
\vspace{0.35cm}
\subfloat[$t_h=0.25$]{
\includegraphics[width=0.95\textwidth]{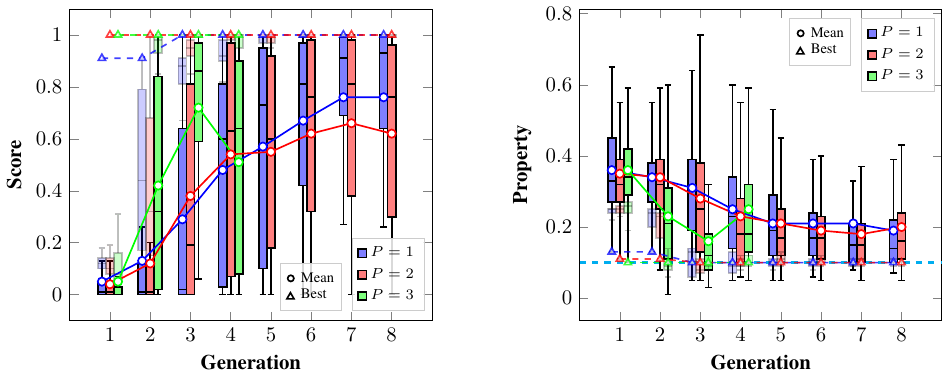}
\label{fig:CsSnI_chemsys_e_0p25}
}
\vspace{0.35cm}
\subfloat[$t_h=0.5$]{
\includegraphics[width=0.95\textwidth]{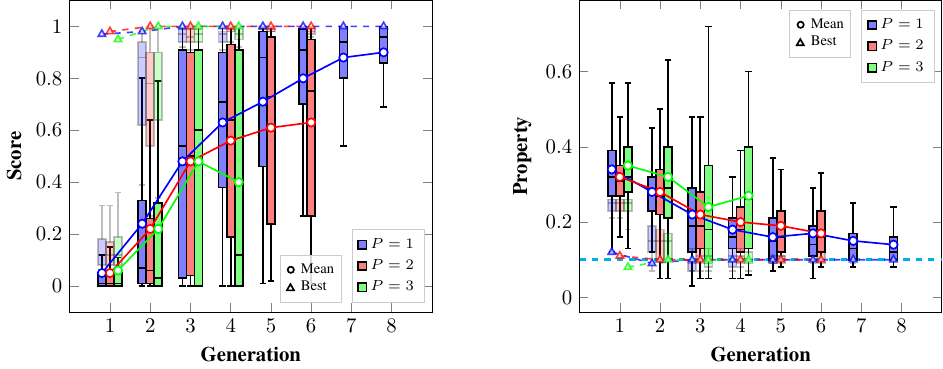}
\label{fig:CsSnI_chemsys_e_0p5}
}
\end{figure}

\begin{figure}[ht!]\ContinuedFloat
\centering
\subfloat[$t_h=0.75$]{
\includegraphics[width=0.95\textwidth]{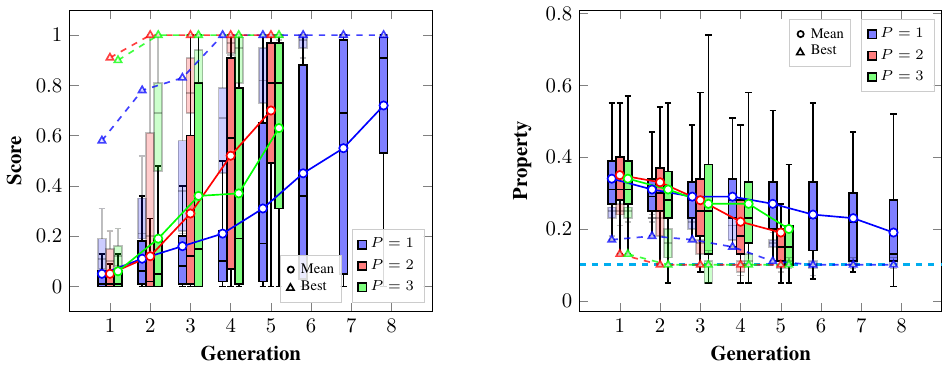}
\label{fig:CsSnI_chemsys_e_0p75}
}
\vspace{0.35cm}
\subfloat[$t_h=0.98$]{
\includegraphics[width=0.95\textwidth]{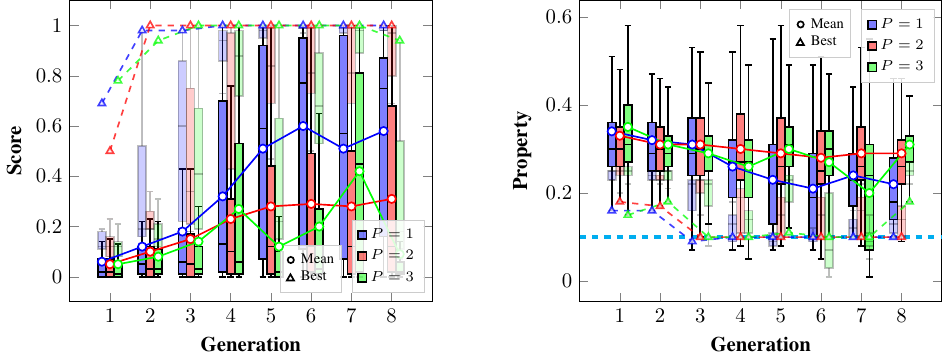}
\label{fig:CsSnI_chemsys_e_0p98}
}
\vspace{0.35cm}
\subfloat[Individual performance across seeds for $t_h=0.002$ and $P=3$ to explain the behavior seen in Fig.~\ref{fig:CsSnI_chemsys_e_0p002}.]{
\includegraphics[width=0.95\textwidth]{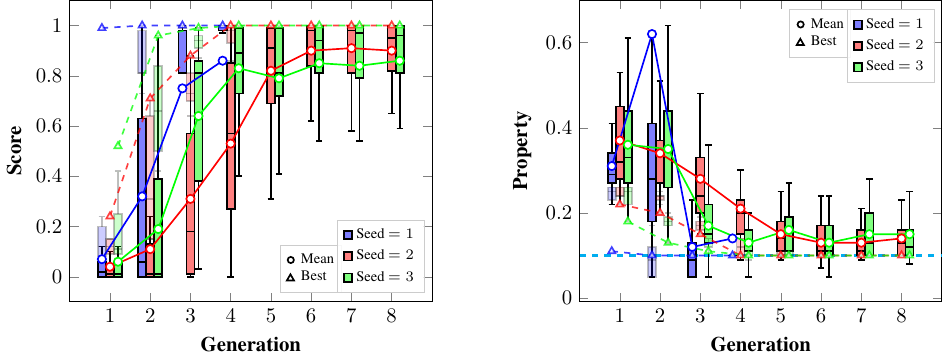}
\label{fig:CsSnI_chemsys_e_0p002_seeds}
}
\caption{Distribution of scores \& properties across generations for experiment B (\texttt{chemical\_system\_energy\_above\_hull} model).}
\label{fig:all_CsSnI_chemsys_e_boxplots}
\end{figure}

\newpage

\begin{figure}[ht!]
\centering
\subfloat[$t_h=0.002$]{
\includegraphics[width=0.95\textwidth]{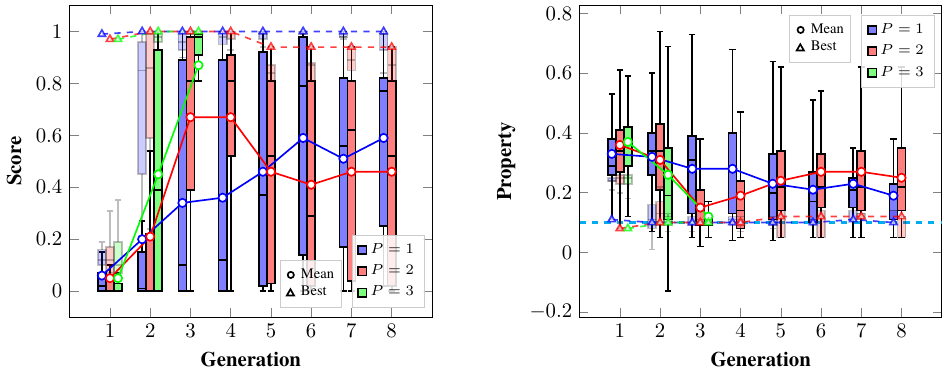}
\label{fig:CsSnI_chemsys_0p002}
}
\vspace{0.35cm}
\subfloat[$t_h=0.25$]{
\includegraphics[width=0.95\textwidth]{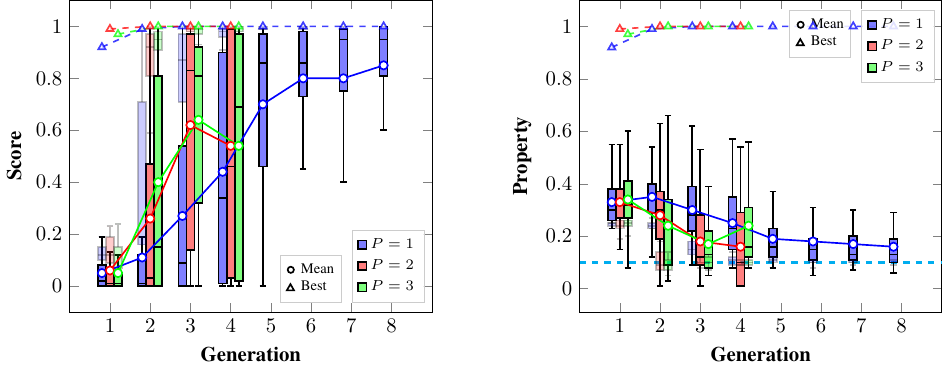}
\label{fig:CsSnI_chemsys_0p25}
}
\vspace{0.35cm}
\subfloat[$t_h=0.5$]{
\includegraphics[width=0.95\textwidth]{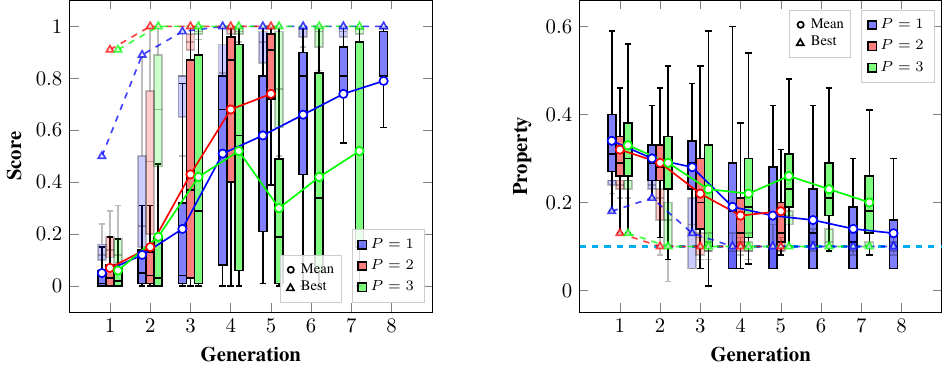}
\label{fig:CsSnI_chemsys_0p5}
}
\end{figure}

\begin{figure}[ht!]\ContinuedFloat
\centering
\subfloat[$t_h=0.75$]{
\includegraphics[width=0.95\textwidth]{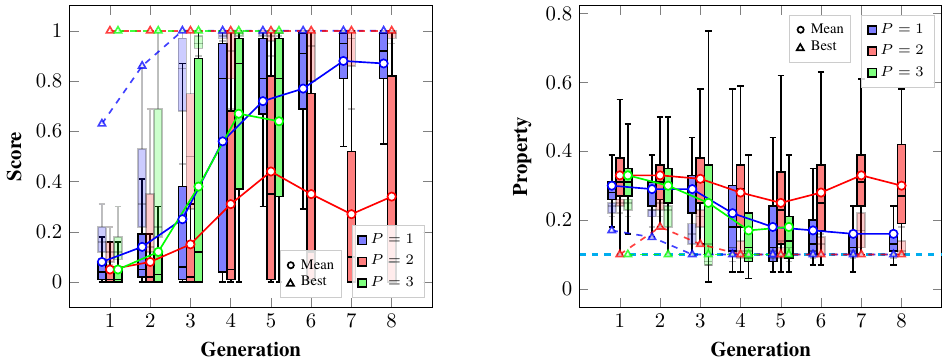}
\label{fig:CsSnI_chemsys_0p75}
}
\vspace{0.35cm}
\subfloat[$t_h=0.98$]{
\includegraphics[width=0.95\textwidth]{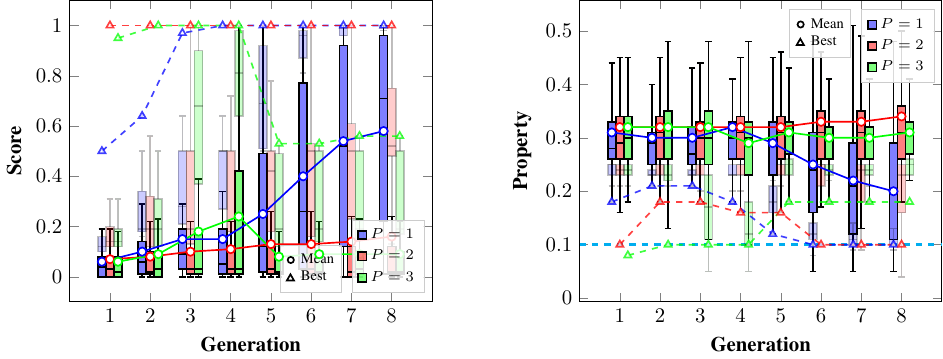}
\label{fig:CsSnI_chemsys_0p98}
}
\vspace{0.35cm}
\subfloat[Individual performance across seeds for $t_h=0.98$ and $P=3$ to explain the behavior seen in Fig.~\ref{fig:CsSnI_chemsys_0p98}.]{
\includegraphics[width=0.95\textwidth]{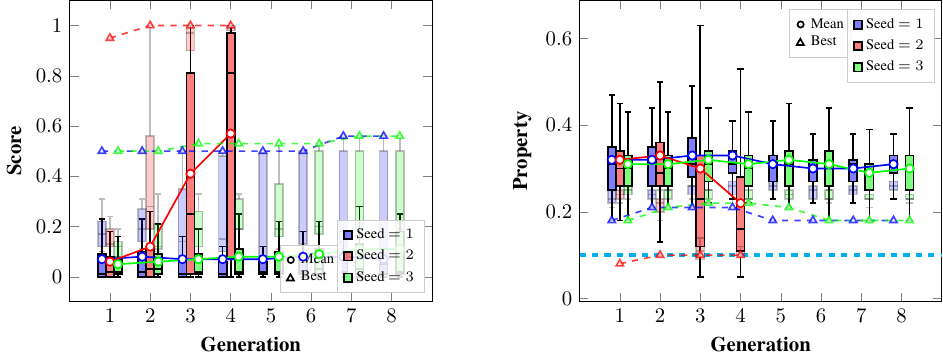}
\label{fig:CsSnI_chemsys_0p98_seeds}
}
\caption{Distribution of scores \& properties across generations for experiment B (\texttt{chemical\_system} model).}
\label{fig:all_CsSnI_chemsys_boxplots}
\end{figure}

\newpage

\begin{figure}[ht!]
\centering
\subfloat[$t_h=0.002$]{
\includegraphics[width=0.95\textwidth]{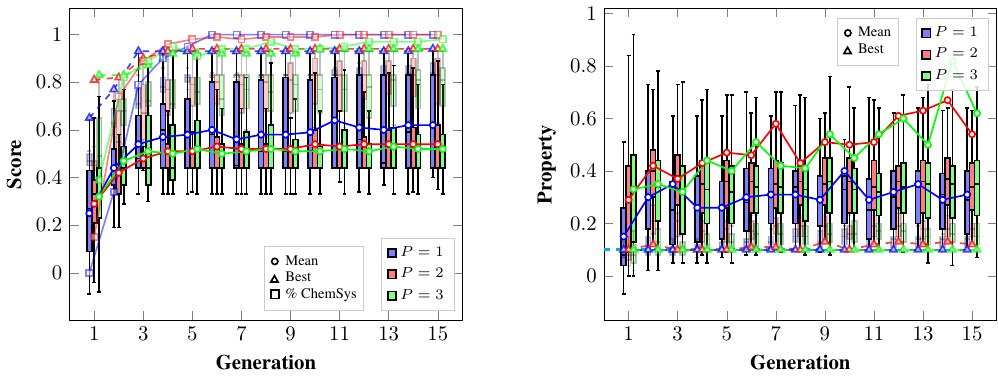}
\label{fig:CsSnI_base_0p002}
}
\vspace{0.35cm}
\subfloat[$t_h=0.25$]{
\includegraphics[width=0.95\textwidth]{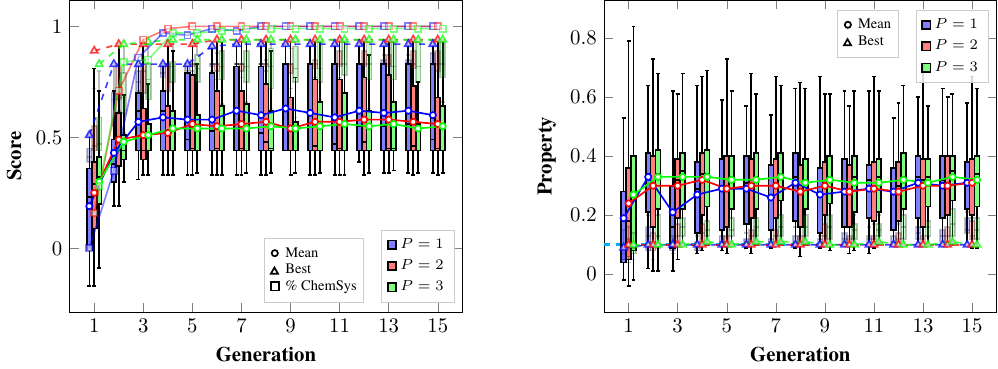}
\label{fig:CsSnI_base_0p25}
}
\vspace{0.35cm}
\subfloat[$t_h=0.5$]{
\includegraphics[width=0.95\textwidth]{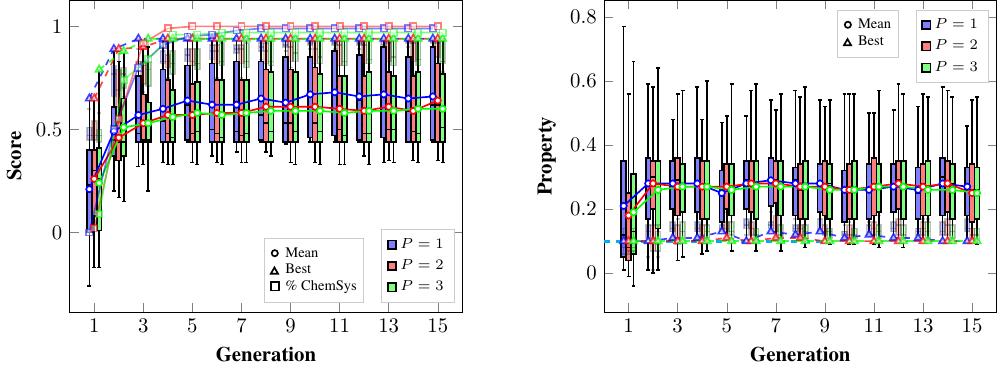}
\label{fig:CsSnI_base_0p5}
}
\end{figure}

\begin{figure}[ht!]\ContinuedFloat
\centering
\subfloat[$t_h=0.75$]{
\includegraphics[width=0.95\textwidth]{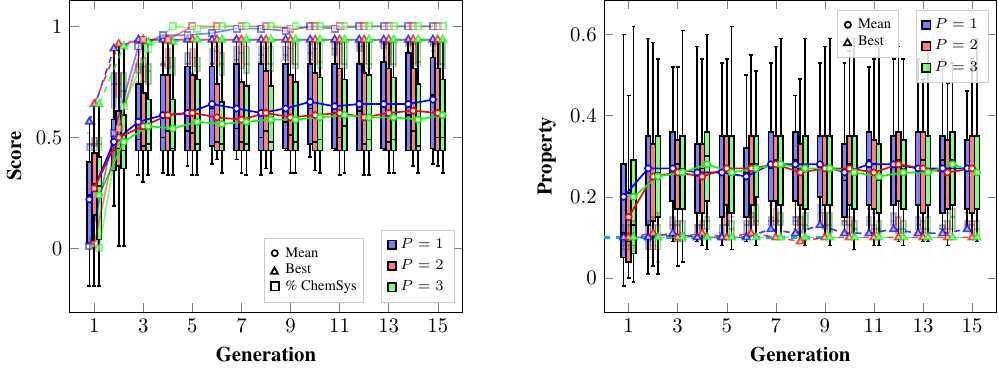}
\label{fig:CsSnI_base_0p75}
}
\vspace{0.35cm}
\subfloat[$t_h=0.98$]{
\includegraphics[width=0.95\textwidth]{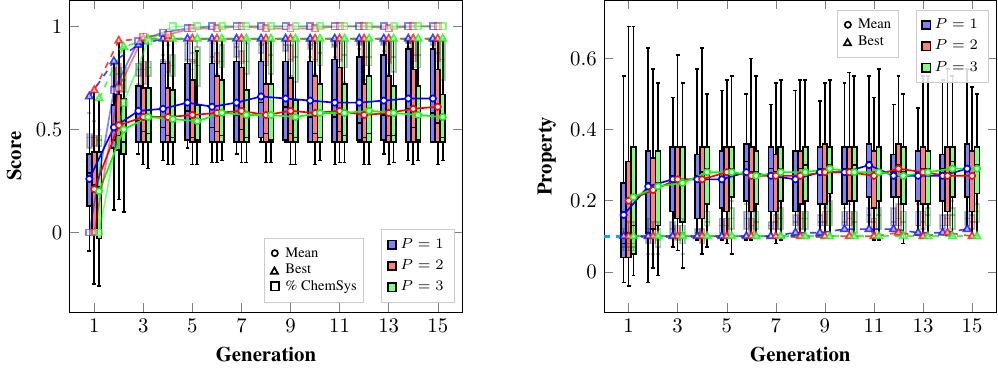}
\label{fig:CsSnI_base_0p98}
}
\vspace{0.35cm}
\subfloat[Individual performance across seeds for $t_h=0.002$ and $P=3$ to explain the behavior seen in Fig.~\ref{fig:CsSnI_base_0p002}.]{
\includegraphics[width=0.95\textwidth]{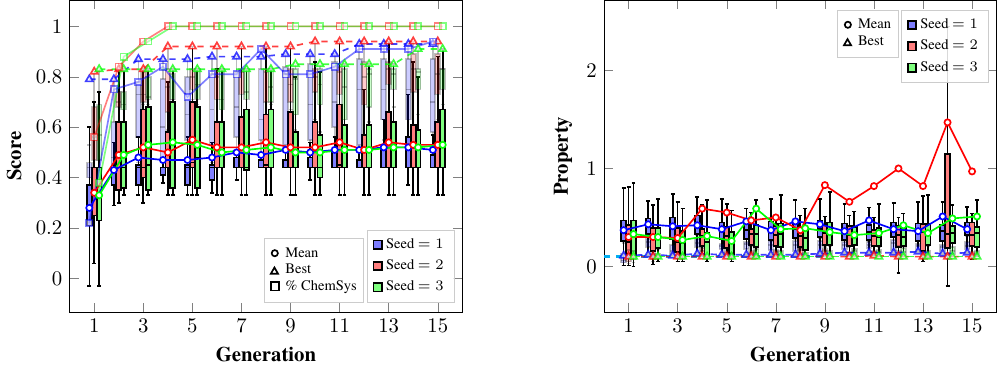}
\label{fig:CsSnI_base_0p002_seeds}
}
\caption{Distribution of scores \& properties across generations for experiment B (\texttt{mattergen\_base} model).}
\label{fig:all_CsSnI_base_boxplots}
\end{figure}

\newpage

\begin{figure}[ht!]
\centering
\subfloat[$t_h=0.002$]{
\includegraphics[width=0.95\textwidth]{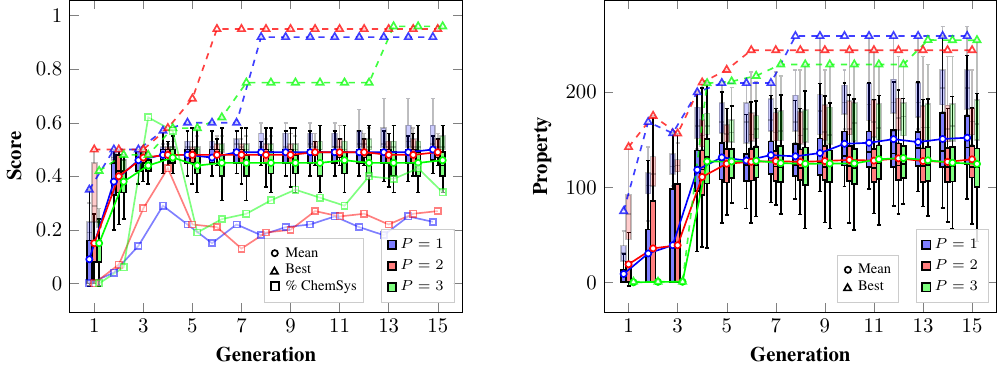}
\label{fig:tmcarbides_base_0p002}
}
\vspace{0.35cm}
\subfloat[$t_h=0.25$]{
\includegraphics[width=0.95\textwidth]{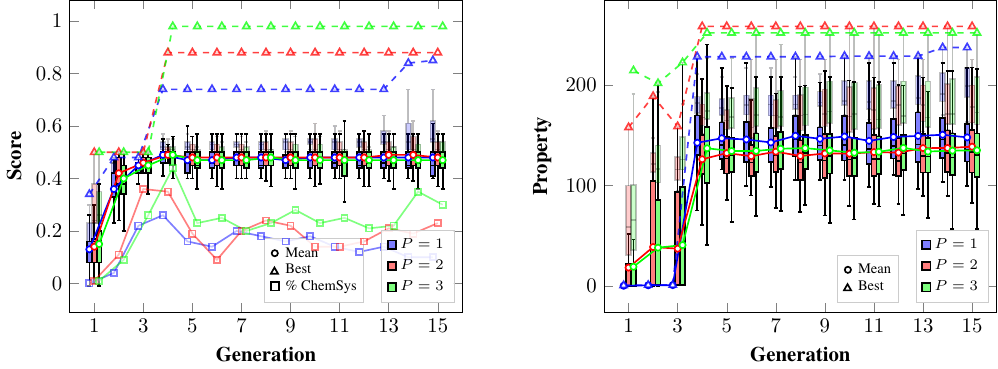}
\label{fig:tmcarbides_base_0p25}
}
\vspace{0.35cm}
\subfloat[$t_h=0.5$]{
\includegraphics[width=0.95\textwidth]{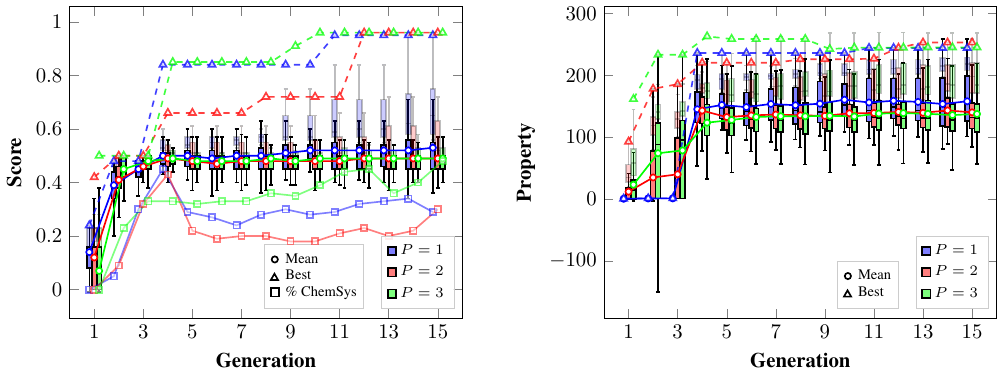}
\label{fig:tmcarbides_base_0p5}
}
\end{figure}

\begin{figure}[ht!]\ContinuedFloat
\centering
\subfloat[$t_h=0.75$]{
\includegraphics[width=0.95\textwidth]{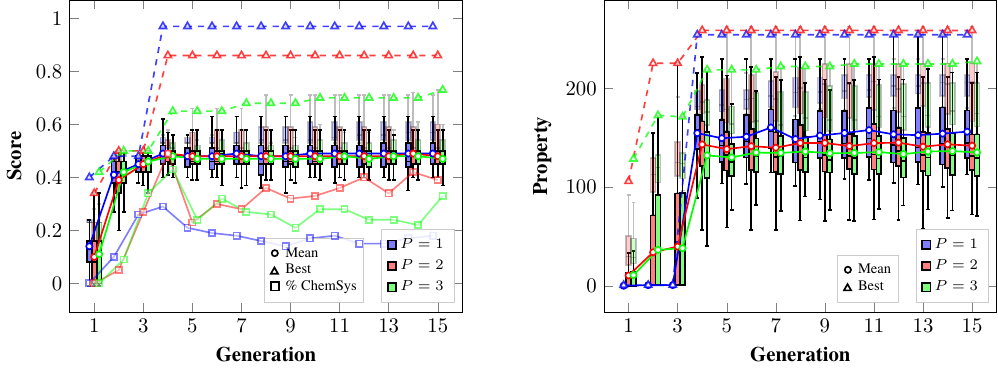}
\label{fig:tmcarbides_base_0p75}
}
\vspace{0.35cm}
\subfloat[$t_h=0.98$]{
\includegraphics[width=0.95\textwidth]{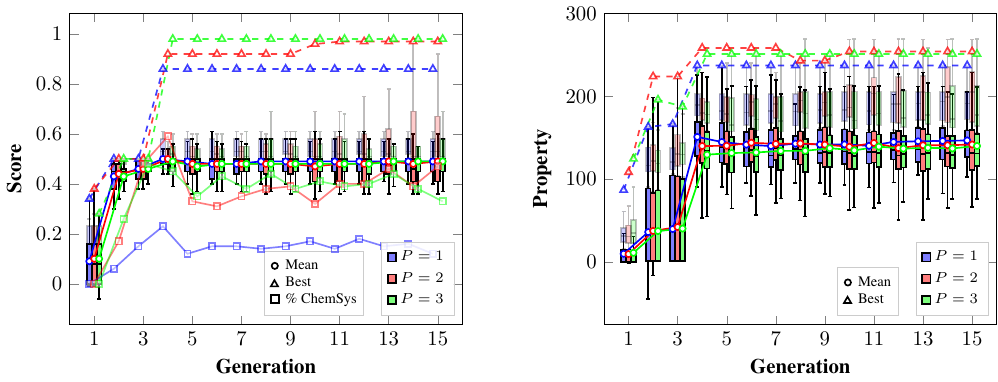}
\label{fig:tmcarbides_base_0p98}
}
\caption{Distribution of scores \& properties across generations for experiment C (\texttt{mattergen\_base} model).}
\label{fig:all_tmcarbides_boxplots}
\end{figure}

\clearpage
\section{Hit rate evolution across generations}
\label{sec:hit_rate_across_gens}
We present the evolution of the number of hits and SUN hits across generations for experiments A and C. These plots complement the results shown in the main manuscript by illustrating how performance develops over time under different objectives and model variants. Consistent with the trends observed for experiment B in Fig. 2 of the main manuscript, DiffGA exhibits a steady increase in both metrics across generations, while GA typically achieves competitive hit counts but lags in SUN-consistent hits, reflecting reduced structural diversity among successful candidates. Together, these figures provide additional insight into the dynamics of evolutionary feedback across varied problem settings and reinforce the generality of the observed behavior.

\begin{figure}[ht!]
\centering
\subfloat[$P=1$]{
\includegraphics[width=0.31\textwidth]{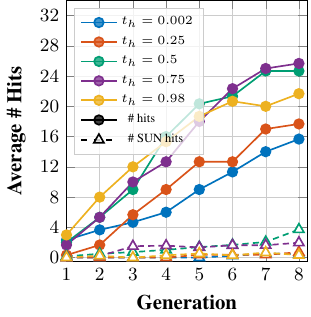}
}
\hfill
\subfloat[$P=2$]{
\includegraphics[width=0.31\textwidth]{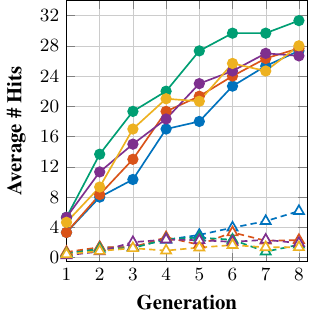}
}
\hfill
\subfloat[$P=3$]{
\includegraphics[width=0.31\textwidth]{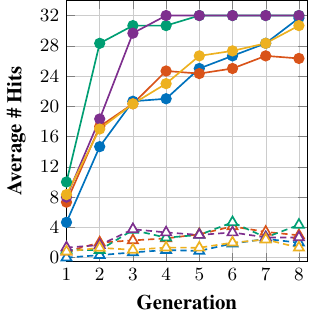}
}
\vspace{0.5cm}
\subfloat[$P=1$]{
\includegraphics[width=0.31\textwidth]{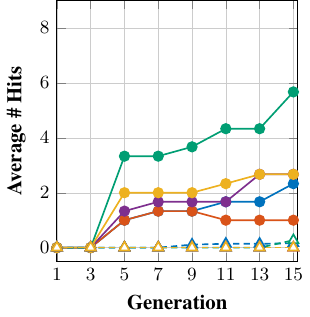}
}
\hfill
\subfloat[$P=2$]{
\includegraphics[width=0.31\textwidth]{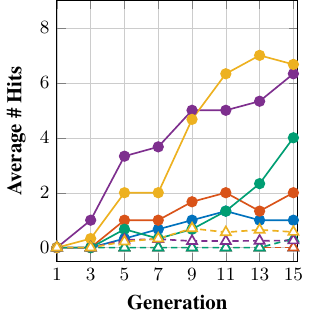}
}
\hfill
\subfloat[$P=3$]{
\includegraphics[width=0.31\textwidth]{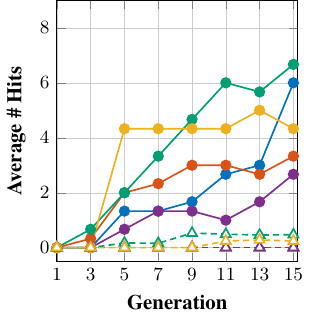}
}
\caption{The average number of hits across generations for (a, b, c) experiment A and (d, e, f) experiment C for each value of $P$, averaged across the three seeds. For clarity, the legend has only been shown in the first plot but applies to all figures, and for (d, e, f), we plot every $2^{\text{nd}}$ data point.}
\label{fig:hits_experiments_A_and_C}
\end{figure}

\newpage
\section{Multi-objective trade-off analysis via Pareto frontiers}
\label{sec:all_results_pareto}
In this section, we provide a multi-objective analysis of DiffGA using Pareto frontiers across experiments A and C (a similar figure for experiment B is presented in the main manuscript). Each frontier captures the trade-off between task success (or property score) and structural quality metrics such as SUN. We note here that in tightly constrained optimization problems, the feasible solution space may be inherently limited, and effective algorithms are expected to concentrate probability mass around a small number of high-quality candidates. The results in this section must be absorbed with trends in analyses presented in Supplementary sections \ref{sec:scores_and_props} and \ref{sec:hit_rate_across_gens}.

\begin{figure}[ht!]
\centering
\subfloat[Experiment A]{
\includegraphics[width=0.45\textwidth]{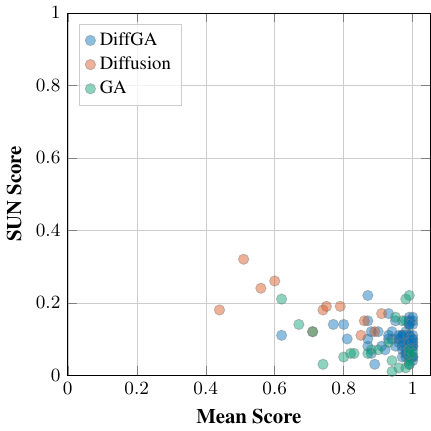}
}
\hfill
\subfloat[Experiment C]{
\includegraphics[width=0.45\textwidth]{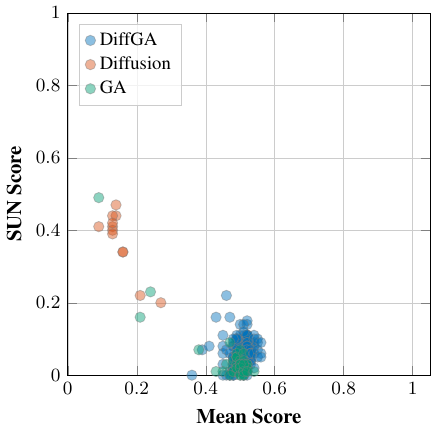}
}
\caption{Pareto frontier-based trade-off analysis for (a) experiment A and (b) experiment C.}
\end{figure}

\end{document}